\documentclass[onecolumn,preprintnumbers,amsmath,amssymb,notitlepage]{revtex4}
\usepackage{graphicx}
\usepackage{epsfig}
\usepackage{amsmath}
\usepackage{amssymb}
\usepackage{dcolumn}

\begin{document}

\title{Quaternionic Quantization Principle in General Relativity and Supergravity}

\author{Martin Kober}
\email{MartinKober@t-online.de}

\affiliation{Kettenhofweg 121, 60325 Frankfurt am Main, Germany}

%\date{\today}

\begin{abstract}
A generalized quantization principle is considered, which incorporates nontrivial commutation relations of
the components of the variables of the quantized theory with the components of the corresponding canonical
conjugated momenta referring to other space-time directions. The corresponding commutation relations are
formulated by using quaternions. At the beginning, this extended quantization concept is applied to the
variables of quantum mechanics. The resulting Dirac equation and the corresponding generalized expression
for plane waves are formulated and some consequences for quantum field theory are considered. Later, the
quaternionic quantization principle is transferred to canonical quantum gravity. Within quantum
geometrodynamics as well as the Ashtekar formalism the generalized algebraic properties of the operators
describing the gravitational observables and the corresponding quantum constraints implied by the
generalized representations of these operators are determined. The generalized algebra also induces
commutation relations of the several components of the quantized variables with each other. Finally,
the quaternionic quantization procedure is also transferred to $\mathcal{N}=1$ supergravity. Accordingly,
the quantization principle has to be generalized to be compatible with Dirac brackets, which appear in
canonical quantum supergravity.
\end{abstract}

\maketitle

\section{Introduction}

The unification of quantum theory with general relativity is probably the most important research topic
in contemporary fundamental theoretical physics. Various approaches exist to obtain a quantum theory
of gravity. In the existing literature one can distinguish between two classes of theories. One class of theories
presupposes usual general relativity, perhaps in a modified formulation, and then transfers the quantization
principle of quantum theory to the corresponding degrees of freedom contained in the gravitational field.
This is performed by canonical quantization or covariant quantization. The other class of theories assumes
a modification of usual general relativity by presupposing an extended geometrical structure of space-time
or an extended dynamics of the gravitational field for example, and then uses the general quantization
principle of quantum theory as well. This means that the quantization principle and thus quantum theory
remains usually completely unchanged.

But in principle it is also thinkable that quantum theory, this means the quantization principle, which
determines the properties of the corresponding quantized theory, if a classical theory is presupposed,
has to be generalized instead of the theory, which shall be formulated quantum theoretically.
This means that it is not only possible to consider the presupposed geometry or dynamics of usual
general relativity as an approximation to a more general gravity theory, but also to consider the
quantization principle usually used in quantum mechanics, quantum field theory and approaches to a
quantum description of general relativity as approximation to a more general quantization principle. 

The generalized uncertainty principle in quantum mechanics as well as noncommutative geometry represent
extensions of the quantum properties of the variables of quantum mechanics.
The generalized uncertainty principle, developed in \cite{Maggiore:1993kv},\cite{Maggiore:1993rv},\cite{Maggiore:1993zu},\cite{Kempf:1994su},\cite{Hinrichsen:1995mf},\cite{Kempf:1996ss},\cite{Kempf:1996nk}, postulates generalized
commutation relations between the position operators and the corresponding momentum operators.
Noncommutative geometry, originally considered in \cite{Snyder:1946qz}, postulates besides the commutation
relations between the position operators and the corresponding momentum operators also commutation relations
between the several components of the position operator and this idea can be transferred to additional
commutation relations between the several momenta, presupposed in \cite{Kober:2010um} for example.
A generalized uncertainty principle can, depending on the special scenario, also imply commutation relations
of the several components of the position operator with each other and this holds analogously for the momentum operator.
These concepts can be interpreted as fundamental properties of nature and thus they would belong to quantum
theory itself and accordingly represent a generalization of the concept of quantization. If this is postulated,
then these generalized quantization principles have also to be transferred to the quantization of
general relativity an thus the gravitational field what differs from the formulation of clasical general
relativity on noncommutative space-time \cite{Calmet:2005qm} or even usual quantum general relativity on
noncommutative space-time \cite{Faizal:2011wm},\cite{Faizal:2012kc},\cite{Faizal:2013ioa},\cite{Kober:2014wsa}.
In \cite{Majumder:2011ad},\cite{Majumder:2011bv},\cite{Majumder:2011eg},\cite{Majumder:2011xg},\cite{Majumder:2011hy}
ideas to transfer the concept of a generalized uncertainty principle to gravity can be found, but in \cite{Kober:2011uj}
and \cite{Garattini:2015aca} the generalized uncertainty principle principle has really been transferred to the
variables of canonical quantum gravity and quantum cosmology, whereas in \cite{Kober:2011am} the concept of
noncommutative geometry has been transferred to the components of the tetrad field. An extension of the field
theoretic quantization principle to a nonlocal quantization principle has been considered in \cite{Kober:2014xxa}.

In the present paper is suggested an approach to generalize the quantization principle of quantum theory,
which seems concerning its application in quantum mechanics as a natural extension of the concept of  
noncommutative geometry. In noncommutative geometry the commutation relations of quantum mechanics are extended
by commutation relations between the several components of the position operator.
In the quantization concept presented in this paper are not only postulated nontrivial commutation relations
between the components of the variables of the theory, which has to be quantized, and the corresponding components
of the canonical conjugated variables belonging to the same space-time direction, but also nontrivial commutation
relations with the components of the canonical conjugated variables belonging to other space-time directions.
This generalized quantization principle of general quantum theory is formulated based on the
mathematical concept of quaternions. Concretely, the additional commutation relations are assumed to be
of the same shape, but are not proportional to the imaginary unit, but to another direction in the space
of quaternions. If all these commutation relations were assumed to be proportional to the usual imaginary
unit of the space of complex numbers, the tensor defining the quantization would not be invertible and
in case of quantum mechanics no plane waves could be defined.

The generalized quantization principle, which could be called as quaternionic quantization principle, is first
studied in the simplest case of quantum mechanics and after this it is transferred to general relativity
and $\mathcal{N}=1$ supergravity. Of course, this kind of generalization of the quantization principle presupposes
canonical quantization and accordingly the canonical formulations of general relativity and $\mathcal{N}=1$
supergravity have to be considered.

Quaternions with respect to physical theories have been studied in many areas of theoretical physics,
for example quaternionic formulations have already been studied in the context of quantum mechanics,  \cite{Finkelstein:1961tk},\cite{Nash:1986rq},\cite{Nash:1987cv},\cite{Adler:1987gd},
\cite{Rotelli:1988fc},\cite{Davies:1990pm},\cite{Horwitz:1992hk},\cite{DeLeo:1993av},\cite{Horwitz:1993bx},\cite{Horwitz:1993by},\cite{DeLeo:1995yq},\cite{Adler:1996qi},\cite{Horwitz:1997ny},\cite{DeLeo:2000ik},\cite{Maia:2001dm},\cite{Schwartz:2006if},\cite{Schwartz:2007ih},\cite{Trifonov:2007wg} and quantum field theory \cite{Adler:1985wz},\cite{Adler:1985uh},\cite{Albeverio:1986vf},\cite{DeLeo:1991mi}.
Concerning general relativity and geometry, quaternionic structures have been studied in
\cite{Edmonds:1974yq},\cite{Lord:1975jy},\cite{Teli:1980mr},\cite{Morita:1983kw},\cite{Page:1985hg},\cite{Razon:1989gm},\cite{Ferrara:1989ik},\cite{deWit:1990na},\cite{deWit:1991nm},
\cite{deWit:1993rr},\cite{Evans:1992az},\cite{Adler:1996tm},\cite{Tao:1996qk},\cite{Kanno:1999hr},\cite{Ivanov:1999vg},\cite{Ivanov:1999sg},\cite{deWit:2001dj},\cite{deWit:2001bk},\cite{Casteill:2001zk},\cite{Hiragane:2003qq},\cite{Mulsch:2003vz},\cite{D'Auria:2004kx},\cite{Andrianopoli:2004im},\cite{Smet:2004da},\cite{Aoyama:2005hb},\cite{Alexandrov:2006hx},\cite{Meusburger:2007ad}.
Besides, quaternions have been used with respect to reformulations and extensions of particle physics and the standard model,
\cite{Adler:1979gv},\cite{Morita:1982af},\cite{Adler:1994pp},\cite{DeLeo:1995jt},\cite{DeLeo:1996af},
\cite{DeLeo:1996ac},\cite{Lim:1997ez},\cite{DeLeo:1998ye},\cite{Scolarici:2000sp},\cite{Andrianopoli:2002vq},\cite{Frigerio:2004jg},\cite{Frigerio:2007nn},\cite{Aranda:2011dx},
and especially with respect to supersymmetry as well as supergravity, \cite{Morita:1984pu},\cite{Adler:1986ck},\cite{Bodner:1989cg},
\cite{D'Auria:1990fj},\cite{Cederwall:1992tg},\cite{Behrndt:2001km},\cite{Toppan:2004xy},\cite{Bellucci:2004pu},\cite{D'Auria:2004tr},\cite{Rawat:2007dd},\cite{Rawat:2007vm}.
But it is very important to mention that the quaternionic generalization of the quantization principle considered
in the present paper differs decisively from earlier considerations, since here a completely new quantization principle
as general physical concept is considered, whereas in the mentioned explorations a quaternionic reformulation of
some physical theories or special geometrical scenarios through the introduction of quaternionic quantities
have been considered. This means that the concept of quaternions serves as a mathematical concept to generalize
the quantization principle as a physical concept, which becomes manifest with respect to quantum mechanics, quantum
field theory as well as the quantization of general relativity and supergravity.

The paper is structured as follows: At the beginning is given a short introduction to the concept of quaternions.
Then the suggested generalized quantization principle, which is based on quaternions and consists in the addition
of nontrivial commutation relations between the components of the variables and the components of the corresponding
canonical conjugated variables belonging to other space-time directions, is formulated for quantum mechanics as special
manifestation. The corresponding free Dirac equation and the generalized plane waves are determined. After this
the corresponding generalized propagator of a scalar field as well as the generalized gauge principle of electrodynamics related to local phase invariance are considered. Subsequently, the main aim of this paper is treated, the generalization
of the canonical quantum description of general relativity by the idea of the quaternionic quantization principle.
Accordingly the generalized commutation relations between the position and momentum operator are transferred to the
variables of quantum geometrodynamics as well as to the variables of the Ashtekar formalism. Based on this, the
corresponding quantum constraints are derived, especially the generalized Wheeler-DeWitt equation. As most intricate
manifestation, the quaternionic quantization principle is also applied to an extension of classical general relativity,
canonical supergravity namely and especially $\mathcal{N}=1$ supergravity. Since in the quantization procedure of
supergravity appear Dirac brackets because of the second class constraints, the quaternionic quantization principle
has to be generalized to be applicable to theories, which are usually quantized by defining Dirac brackets. This
leads to much more complicated commutation relations. After this, the corresponding constraints are generalized
and finally the inner product of canonical quantum supergravity has also to be reformulated.

\section{Quaternions}

In this section is given a short repetition of the concept of quaternions, which serves also to introduce the notation,
which is used to express the quaternions. Quaternions are a generalization of complex numbers with two additional
dimensions besides the usual real dimension and the usual imaginary dimension. This means that quaternions represent
elements of a four-dimensional vector space as number space. A quaternion can be represented in the following way:

\begin{equation}
q=a+bi+cj+dk,
\label{quaternion}
\end{equation}
where $a$, $b$, $c$ and $d$ are real numbers and $i$, $j$, $k$ are quantities fulfilling the following relations:
       
\begin{eqnarray}
&&ij=-ji=k,\quad jk=-kj=i,\quad ki=-ik=j,\quad\quad\quad\quad
i^2=j^2=k^2=-1.
\label{properties_quaternionic_units}
\end{eqnarray}                                           
The corresponding conjugated quantity to a quaternion $q$ defined in ($\ref{quaternion}$), $q^{*}$, is defined as

\begin{equation}
q^{*}=a-bi-cj-dk.
\label{quaternionic_conjugation}
\end{equation}
Of course, the space of the complex numbers represents a subspace of the space of the quaternions, which is built by
all quaternions with $c=d=0$. The norm of a quaternion denoted by $|q|$ is given in analogy to the norm of a complex
number by

\begin{equation}
|q|=\sqrt{q^{*}q}=\sqrt{a^2+b^2+c^2+d^2}.
\end{equation}
The quaternions can be represented by using the Pauli matrices, if $1$, $i$, $j$ and $k$ are related to the
Pauli matrices the unity matrix in two dimensions included in the following way,

\begin{equation}
{\bf 1}=\sigma^0=\left(\begin{matrix}1 & 0\\0 & 1 \end{matrix}\right),\quad
i=i_{\mathbb{C}}\sigma^3=\left(\begin{matrix}i_{\mathbb{C}} & 0 \\0 & -i_{\mathbb{C}} \end{matrix}\right),\quad
j=i_{\mathbb{C}}\sigma^2=\left(\begin{matrix}0 & 1\\-1 & 0\end{matrix}\right),\quad
k=i_{\mathbb{C}}\sigma^1=\left(\begin{matrix}0 & i_{\mathbb{C}}\\i_{\mathbb{C}} & 0 \end{matrix}\right),
\end{equation}
where $i_{\mathbb{C}}$ denotes the usual complex unit. This is helpful to determine the inverse matrix
of a quaternionic matrix for example, which can be represented by a complex matrix in this way.

\section{Quaternionic Quantization in Quantum Mechanics}

In this section the generalized quantization concept based on the mathematical concept of quaternions
shall be introduced and this is done in the realm of quantum mechanics. The basic postulate of quantum
mechanics consists in the fundamental commutation relation between the position and the corresponding
momentum operators,

\begin{equation}
\left[\hat x^\mu, \hat p_\nu \right]=i\delta^{\mu}_{\nu},
\label{Usual_Quantization_QM}
\end{equation}
which defines their mathematical properties and constitutes the corresponding complex vector space
of the possible states. The Planckian constant $\hbar$ as well as the speed of light $c$ are set
equal to one throughout the paper, $\hbar=c=1$. Greek indices refer to all coordinates of space-time, whereas
Latin indices refer to the spatial coordinates of space-time in this paper. In usual quantum mechanics the
components of the position operator fulfil only nontrivial commutation relations with the corresponding
components of the momentum operator, which refer to the same space-time direction.
Noncommutative geometry extends these relations by additional commutation relations between the
several components of the space-time coordinates, $\left[\hat x^\mu,\hat x^\nu\right]=i\theta^{\mu\nu}$. 
As already mentioned in the explanation of the introduction, it suggests itself to generalized the
quantization postulate ($\ref{Usual_Quantization_QM}$) to a generalized quantization postulate,
containing also nontrivial commutation relations between the components of the position operator
and the components of the momentum operator not referring to the same space-time direction.
If these commutation relations would also be postulated to be proportional to the complex imaginary
unit $i_{\mathbb{C}}$, this would lead to a matrix, which is not invertible and thus no generalized
plane waves as solutions for the free field equations could be defined. Therefore quaternions are
introduced to enable the postulation of commutation relations being proportional to another
direction in the space of the quaternions. Accordingly, a transition to the following fundamental
commutation relation is suggested as generalization of usual quantum mechanics:

\begin{equation}
\left[\hat x^\mu, \hat p_\nu \right]=i\delta^{\mu}_{\nu}\quad\longrightarrow\quad
\left[\hat x^\mu, \hat p_\nu \right]=\alpha^{\mu}_{\ \nu},
\label{Quaternionic_Quantization_QM}
\end{equation}
where $\alpha^{\mu}_{\ \nu}$ is a quaternionic tensor of second order which looks as follows:

\begin{equation}
\alpha^{\mu}_{\ \nu}=\left(\begin{matrix}i & \varkappa j & \varkappa j & \varkappa j\\
\varkappa j & i & \varkappa j & \varkappa j \\
\varkappa j & \varkappa j &  i & \varkappa j\\
\varkappa j &  \varkappa j & \varkappa j & i \end{matrix}\right).
\label{quaternionic_tensor_QM}
\end{equation}
$i$ and $j$ denote the units of the quaternionic number space defined in ($\ref{properties_quaternionic_units}$) and
$\varkappa$ denotes a dimensionless parameter defining the relation between the influence of the usual commutation
relations and the additional ones. If $\varkappa$ goes to zero, one obtains usual quantum mechanics as approximation
to the theory presented in this paper. To generalize plane waves according to ($\ref{Quaternionic_Quantization_QM}$),
the inverse matrix of $\alpha^{\mu}_{\ \nu}$ will become important, which is given by 

\begin{equation}
\left(\alpha^{-1}\right)^{\mu}_{\ \nu}=
\left(\begin{matrix}
a_i i+a_j j & b_i i+b_j j & b_i i+b_j j & b_i i+b_j j \\
b_i i+b_j j & a_i i+a_j j & b_i i+b_j j & b_i i+b_j j \\
b_i i+b_j j & b_i i+b_j j & a_i i+a_j j & b_i i+b_j j \\
b_i i+b_j j & b_i i+b_j j & b_i i+b_j j & a_i i+a_j j \\
\end{matrix}\right),
\end{equation}
where the coefficients $a_i$, $a_j$, $b_i$ and $b_j$ are defined as

\begin{eqnarray}
a_i=-\frac{7\varkappa^2+1}{9\varkappa^4+10\varkappa^2+1},\quad
a_j=\frac{6\varkappa^3}{9\varkappa^4+10\varkappa^2+1},\quad
b_i=\frac{2\varkappa^2}{9\varkappa^4+10\varkappa^2+1},\quad
b_j=-\frac{3\varkappa^3+\varkappa}{9\varkappa^4+10\varkappa^2+1}.
\label{coefficients_inverse_alpha}
\end{eqnarray}
The quantization postulate ($\ref{Quaternionic_Quantization_QM}$) also implies nontrivial commutation relations
between the several components of the position and the momentum operator, which are implied by the nontrivial
commutation relations between the several components of the tensor $\alpha^{\mu}_{\ \nu}$, which are given by

\begin{equation}
\left[\alpha^{\mu\nu},\alpha^{\rho\sigma}\right]=\Lambda^{\mu\nu\rho\sigma},
\label{commutator_quantization_tensor_QM}
\end{equation}
where $\Lambda^{\mu\nu\rho\sigma}$ is a tensor of fourth order,
which can be represented as a matrix,

\begin{equation}
\Lambda^{\mu\nu\rho\sigma}=\varkappa\left(\begin{matrix}\lambda_i^{\mu\nu} & \lambda_j^{\mu\nu}
& \lambda_j^{\mu\nu} & \lambda_j^{\mu\nu}\\
\lambda_j^{\mu\nu} & \lambda_i^{\mu\nu} & \lambda_j^{\mu\nu} & \lambda_j^{\mu\nu}\\
\lambda_j^{\mu\nu} & \lambda_j^{\mu\nu} & \lambda_i^{\mu\nu} & \lambda_j^{\mu\nu}\\
\lambda_j^{\mu\nu} & \lambda_j^{\mu\nu} & \lambda_j^{\mu\nu} & \lambda_i^{\mu\nu}\end{matrix}\right),
\end{equation}
containing the tensors $\lambda_i^{\mu\nu}$ and $\lambda_j^{\mu\nu}$ of second order, which are of
the following shape, if they are again represented as matrices:

\begin{eqnarray}
\lambda_i^{\mu\nu}=\left(\begin{matrix}0 & 2k & 2k & 2k\\ 2k & 0  & 2k & 2k\\
2k & 2k & 0 & 2k\\ 2k & 2k & 2k & 0 \end{matrix}\right),\quad
\lambda_j^{\mu\nu}=\left(\begin{matrix}-2k & 0 & 0 & 0\\ 0 & -2k & 0 & 0\\
0 & 0 & -2k & 0\\ 0 & 0 & 0 & -2k
\end{matrix}\right),
\end{eqnarray}
where $k$ denotes according to ($\ref{properties_quaternionic_units}$) besides $i$ and $j$ the third imaginary
unit in the quaternionic number space.
The generalized quantization condition ($\ref{Quaternionic_Quantization_QM}$) is fulfilled by the following
shape of the position and the momentum operator represented in position space:

\begin{equation}
\hat x^\mu=x^\mu,\quad \hat p_\mu=-\alpha^{\nu}_{\ \mu}\frac{\partial}{\partial x^\nu}.
\label{position_representation_operators_QM}
\end{equation}
%If the components of the momentum operator $\hat p_\mu$ represented in position space, which is given in
%($\ref{position_representation_operators_QM}$), are written explicitly, they look as follows:

%\begin{eqnarray}
%\hat p_0&=&-i\frac{\partial}{\partial x_0}-\varkappa j\frac{\partial}{\partial x_1}
%-\varkappa j\frac{\partial}{\partial x_2}-\varkappa j\frac{\partial}{\partial x_3},\nonumber\\
%\hat p_1&=&-\varkappa j\frac{\partial}{\partial x_0}-i\frac{\partial}{\partial x_1}
%-\varkappa j\frac{\partial}{\partial x_2}-\varkappa j\frac{\partial}{\partial x_3},\nonumber\\
%\hat p_2&=&-\varkappa j\frac{\partial}{\partial x_0}-\varkappa j\frac{\partial}{\partial x_1}
%-i\frac{\partial}{\partial x_2}-\varkappa j\frac{\partial}{\partial x_3},\nonumber\\
%\hat p_3&=&-\varkappa j\frac{\partial}{\partial x_0}-\varkappa j\frac{\partial}{\partial x_1}
%-\varkappa j\frac{\partial}{\partial x_2}-i\frac{\partial}{\partial x_3}.
%\end{eqnarray}
The position and the momentum operator represented in momentum space are of
the following shape:

\begin{equation}
\hat x^\mu=\alpha^\mu_{\ \nu}\frac{\partial}{\partial p_\nu},\quad \hat p^\mu=p^\mu.
\label{momentum_representation_operators_QM}
\end{equation}
By using the representation of the momentum operator in position space given in ($\ref{position_representation_operators_QM}$)
and the representation of the position operator in momentum operator given in ($\ref{momentum_representation_operators_QM}$),
the commutation relations of the several components of the momentum operator with each other and of the several components of
the position operator with each other can be determined. The position representation is only valid, if the algebra ($\ref{Quaternionic_Quantization_QM}$) is supplemented by the following commutation relations:

\begin{eqnarray}
[\hat x^\mu,\hat x^\nu]=0,\quad[\hat p^\mu,\hat p^\nu]=[\alpha^{\mu\rho},\alpha^{\nu\sigma}]\frac{\partial}{\partial x^\rho}\frac{\partial}{\partial x^\sigma}
=\Lambda^{\mu\rho\nu\sigma}\frac{\partial}{\partial x^\rho}\frac{\partial}{\partial x^\sigma}
=\Lambda^{\mu\rho\nu\sigma}\left(\alpha^{-1}\right)_{\rho\lambda}
\left(\alpha^{-1}\right)_{\sigma\kappa}\hat p^{\lambda}\hat p^{\kappa},
\label{commutators_position_representation}
\end{eqnarray}
and analogously, the momentum representation is only valid, if the algebra ($\ref{Quaternionic_Quantization_QM}$) is supplemented by the following commutation relations:

\begin{eqnarray}
[\hat x^\mu,\hat x^\nu]=[\alpha^{\mu\rho},\alpha^{\nu\sigma}]
\frac{\partial}{\partial p^\rho}\frac{\partial}{\partial p^\sigma}
=\Lambda^{\mu\rho\nu\sigma}\frac{\partial}{\partial p^\rho}\frac{\partial}{\partial p^\sigma}
=\Lambda^{\mu\rho\nu\sigma}\left(\alpha^{-1}\right)_{\rho\lambda}
\left(\alpha^{-1}\right)_{\sigma\kappa}\hat x^{\lambda}\hat x^{\kappa},\quad [\hat p^\mu,\hat p^\nu]=0.
\label{commutators_momentum_representation}
\end{eqnarray}
In ($\ref{commutators_position_representation}$) and ($\ref{commutators_momentum_representation}$), the commutation
relations between the several components of the quantization tensor $\alpha^\mu_{\ \nu}$,
($\ref{commutator_quantization_tensor_QM}$), has been used. The squared four-momentum operator as well as the squared
four-position operator are equal to the corresponding squared operators of usual quantum mechanics and thus the
following commutation relations are valid:

\begin{equation}
\left[\hat p^\mu \hat p_\mu, \hat p_\nu\right]=0,\quad \left[\hat x^\mu \hat x_\mu, \hat x_\nu\right]=0.
\end{equation}
Of course, the commutation relations between the components of the corresponding angular momentum take a
generalized form as well. The angular momentum operator is defined as

\begin{equation}
\hat L^a=\frac{1}{2}\epsilon^{abc}\left(\hat x_b \hat p_c+\hat p_c \hat x_b\right),
\end{equation}
where $\epsilon^{abc}$ denotes the total antisymmetric tensor in three dimensions.
By using ($\ref{Quaternionic_Quantization_QM}$) as well as ($\ref{commutators_position_representation}$) or
($\ref{commutators_momentum_representation}$) respectively, the commutators between the components of the
angular momentum operator with each other in case of the position representation as well as the
momentum representation can be calculated and are given by

\begin{eqnarray}
\left[\hat L^a, \hat L^d\right]=&&\frac{1}{4}\epsilon^{abc}\epsilon^{def}\left[
2 \alpha_{bf} \left(\hat x_e \hat p_c+\hat p_c \hat x_e\right)
-2 \alpha_{ec} \left(\hat x_b \hat p_f+\hat p_f \hat x_b\right)
%\right.\nonumber\\&&\left.
+\Lambda_{c g f h}\left(\alpha^{-1}\right)^{g i}\left(\alpha^{-1}\right)^{h j}
\hat x_b \hat x_e \hat p_{i}\hat p_{j}
\right.\nonumber\\&&\left.
+\Lambda_{c g f h}\left(\alpha^{-1}\right)^{g i}\left(\alpha^{-1}\right)^{h j}
\hat x_b \hat p_{i} \hat p_{j} \hat x_e
%\right.\nonumber\\&&\left.
+\Lambda_{c g f h}\left(\alpha^{-1}\right)^{g i}\left(\alpha^{-1}\right)^{h j}
\hat x_e \hat p_{i} \hat p_{j} \hat x_b
%\right.\nonumber\\&&\left.
+\Lambda_{c g f h}\left(\alpha^{-1}\right)^{g i}\left(\alpha^{-1}\right)^{h j}
\hat p_{i} \hat p_{j} \hat x_b \hat x_e\right]\nonumber\\
\end{eqnarray}
in case of ($\ref{commutators_position_representation}$) corresponding to the position representation, and by

\begin{eqnarray}
\left[\hat L^a, \hat L^d\right]=&&\frac{1}{4}\epsilon^{abc}\epsilon^{def}\left[
2 \alpha_{bf} \left(\hat x_e \hat p_c+\hat p_c \hat x_e\right)
-2 \alpha_{ec} \left(\hat x_b \hat p_f+\hat p_f \hat x_b\right)
%\right.\nonumber\\&&\left.
+\Lambda_{b g e h}\left(\alpha^{-1}\right)^{g i}\left(\alpha^{-1}\right)^{h j}\hat x_{i}\hat x_{j} \hat p_c \hat p_f
\right.\nonumber\\&&\left.
+\Lambda_{b g e h}\left(\alpha^{-1}\right)^{g i}
\left(\alpha^{-1}\right)^{h j}\hat p_f \hat x_{i} \hat x_{j} \hat p_c
%\right.\nonumber\\&&\left.
+\Lambda_{b g e h}\left(\alpha^{-1}\right)^{g i}\left(\alpha^{-1}\right)^{h j}
\hat p_c \hat x_{i} \hat x_{j} \hat p_f
%\right.\nonumber\\&&\left.
+\Lambda_{b g e h}\left(\alpha^{-1}\right)^{g i}\left(\alpha^{-1}\right)^{h j}
\hat p_c \hat p_f\hat x_{i}\hat x_{j}\right]\nonumber\\
\end{eqnarray}
in case of ($\ref{commutators_momentum_representation}$) corresponding to the momentum representation.
The states $|\psi \rangle$ of the Hilbert space $\mathcal{H}_{Q}$, where the operators of the quaternionic
generalized quantum mechanics according to ($\ref{Quaternionic_Quantization_QM}$) live in, can be
represented as wave-functions, which take quaternionic values,

\begin{equation}
\psi(x)=\psi_{\bf 1}(x){\bf 1}+\psi_i(x)i+\psi_j(x)j+\psi_k(x)k.
\label{quaternionic_wave_function}
\end{equation} 
The inner product between two states constituting the Hilbert space $\mathcal{H}_{Q}$,
$\langle\ \cdot\ |\ \cdot\ \rangle$, represented in position space, can be defined
in complete analogy to usual quantum mechanics to be

\begin{equation}
\langle \varphi|\psi \rangle=\int d^3 x\ \varphi^{*}(x)\psi(x),
\label{quaternionic_inner_product_QM}
\end{equation}
where $\psi^{*}(x)$ denotes the conjugated quaternionic wave function, which is defined by 
($\ref{quaternionic_wave_function}$) and the definition of quaternionic conjugation
($\ref{quaternionic_conjugation}$). Also hermitian conjugation of operators is defined in analogy
to usual quantum mechanics by replacing complex conjugation by quaternionic conjugation.
Accordingly the hermitian conjugated operator to an operator $\hat A$ is defined as

\begin{equation}
\hat A^{\dagger}=\hat A^{*T},
\label{quaternionic_adjungation}
\end{equation}
where the $*$ again denotes quaternionic conjugation as specified in ($\ref{quaternionic_conjugation}$)
and generalized hermiticity and unitarity are accordingly defined by the conditions $A^{\dagger}=A$
and $U^{\dagger}=U^{-1}$. The generalized position and the generalized momentum operators are still
hermitian operators. To show this, as usual one uses the relation
$\langle \varphi|\hat A|\psi \rangle=\langle \psi|\hat A^{\dagger}|\varphi \rangle^{*}$ with respect
to the quaternionic case referring to ($\ref{quaternionic_inner_product_QM}$) and
($\ref{quaternionic_adjungation}$), which implies for hermitian operators:
$\langle \varphi|\hat A|\psi \rangle=\langle \psi|\hat A|\varphi \rangle^{*}$.
Concerning the position operator this means

\begin{eqnarray}
\langle \varphi|\hat p_\mu|\psi \rangle&=&\int d^3 x\ \varphi^{*}(x)\left(-\alpha^\nu_{\ \mu}
\frac{\partial}{\partial x^\nu}\right)\psi(x)
=\int d^3 x\ \psi(x) \left(\alpha^\nu_{\ \mu}\frac{\partial}{\partial x^\nu}\right) \varphi^{*}(x)\nonumber\\
&=&\int d^3 x\ \left[\psi^{*}(x)\left(-\alpha^\nu_{\ \mu}
\frac{\partial}{\partial x^\nu}\right)\varphi(x)\right]^{*}
=\langle \psi|\hat p_\mu|\varphi \rangle^{*}.
\label{hermiticity_momentum_operator}
\end{eqnarray}
In ($\ref{hermiticity_momentum_operator}$) has been performed partial integration in the second step,
where has been used that any wave-function $\psi(x)$ representing a physical state $|\psi\rangle$ has
to go to zero at infinity to maintain that the function is square integrable $\int d^3 x |\psi|^2 < \infty$,
and that $\left(\alpha^{*}\right)^{\nu}_{\ \ \mu}=-\alpha^\nu_{\ \mu}$. The hermiticity of the position
operator can of course be shown analogously.
The generalized Klein-Gordon equation corresponding to the generalized representation of the
momentum operators ($\ref{position_representation_operators_QM}$) looks as follows:

\begin{equation}
\left(\hat p^\mu \hat p_\mu-m^2\right)\psi(x)=0
\quad\Leftrightarrow\quad
\left(\alpha^{\mu\nu}\alpha_{\mu\rho}\partial_\nu \partial^\rho-m^2\right)\psi(x)=0,
\label{quaternionic_Klein-Gordon_equation}
\end{equation}
and the corresponding generalized Dirac equation looks as follows:

\begin{equation}
\left(\gamma^\mu \hat p_\mu+m\right)\psi=0\quad\Leftrightarrow\quad
\left(\alpha^{\mu\nu}\gamma_\mu \partial_\nu+m\right)\psi=0.
\label{quaternionic_Dirac_equation}
\end{equation}
The corresponding Dirac Lagrangian to ($\ref{quaternionic_Dirac_equation}$) is of the following form: 

\begin{equation}
\mathcal{L}=\bar \psi \left(\alpha^{\mu\nu}\gamma_\mu \partial_\nu-m\right)\psi,
\end{equation}
where $\bar \psi=\psi^{\dagger}\gamma^0$ and the $\dagger$ denotes quaternionic adjungation
according to ($\ref{quaternionic_adjungation}$).
It is important to mention that the Dirac matrices denoted by $\gamma^\mu$ referring to the
Dirac spinor space are still formulated with usual complex numbers and accordingly they
commute with the quaternionic quantization tensor,

\begin{equation}
\left[\alpha^{\mu\nu},\gamma^\rho\right]=0.
\label{commutator_alpha_gamma}
\end{equation}
The solution of the generalized Dirac equation ($\ref{quaternionic_Dirac_equation}$) as well as
of the corresponding generalized Klein-Gordon equation ($\ref{quaternionic_Klein-Gordon_equation}$)
is defined by the generalized plane waves, which are equivalent to the eigenstates of the momentum
operator in position space, $|p\rangle$, which are given by

\begin{equation}
|p\rangle=\exp\left[-\left(\alpha^{-1}\right)^{\mu\nu}p_\mu x_\nu\right],
\label{momentum_eigenstates_A}
\end{equation}
what can be seen by applying the momentum operator to $|p\rangle$:

\begin{eqnarray}
\hat p_{\mu}|p\rangle&=&-\alpha^\nu_{\ \mu} \partial_\nu \exp\left[-\left(\alpha^{-1}\right)^{\rho\sigma} p_\rho x_\sigma\right]
=-\alpha^\nu_{\ \mu}\partial_\nu\left[\left(-\alpha^{-1}\right)^{\rho\sigma} p_\rho x_\sigma\right]
\exp\left[-\left(\alpha^{-1}\right)^{\rho\sigma} p_\rho x_\sigma\right]\nonumber\\
&=&\alpha^\nu_{\ \mu}\left(\alpha^{-1}\right)_{\ \nu}^{\rho} p_\rho
\exp\left[-\left(\alpha^{-1}\right)^{\rho\sigma} p_\rho x_\sigma\right]
=\delta_\mu^\rho p_\rho \exp\left[-\left(\alpha^{-1}\right)^{\rho\sigma} p_\rho x_\sigma\right]\nonumber\\
&=&p_\mu \exp\left[-\left(\alpha^{-1}\right)^{\rho\sigma} p_\rho x_\sigma\right]=p_\mu |p\rangle.
\label{eigenvalue_equation_momentum}
\end{eqnarray}
To separate the components belonging to $i$ and to $j$, the momentum eigenstates
($\ref{momentum_eigenstates_A}$) can be rewritten to

\begin{eqnarray}
\exp\left[-\left(\alpha^{-1}\right)^{\mu\nu}p_\mu x_\nu\right]
&=&\exp\left[-\left(a_i i p^\mu x_\mu+a_j j p^\mu x_\mu
+b_i i p_0 x_1+b_j j p_0 x_1+b_i ip_0 x_2+b_j j p_0 x_2+b_i ip_0 x_3+b_j j p_0 x_3
\right.\right.\nonumber\\ &&\left.\left.
\quad\quad\quad
+b_i ip_1 x_0+b_j j p_1 x_0+b_i ip_1 x_2+b_j p_1 x_2+b_i i p_1 x_3+b_j j p_1 x_3
\right.\right.\nonumber\\ &&\left.\left.
\quad\quad\quad
+b_i i p_2 x_0+b_j j p_2 x_0+b_i i p_2 x_1+b_j j p_2 x_1+b_i ip_2 x_3+b_j j p_2 x_3
\right.\right.\nonumber\\ &&\left.\left.
\quad\quad\quad
+b_i i p_3 x_0+b_j j p_3 x_0+b_i ip_3 x_1+b_j j p_3 x_1+b_i ip_3 x_2+b_j j p_3 x_2
\right)\right]\nonumber\\
&=&\exp\left[-\left(i\beta_i^{\mu\nu}p_\mu x_\nu+j\beta_j^{\mu\nu}p_\mu x_\nu
\right)\right],
\label{momentum_eigenstates_B}
\end{eqnarray}
where the tensors $\beta_i^{\mu\nu}$ as well as $\beta_j^{\mu\nu}$ have been defined,
which can be represented as matrices in the following way:

\begin{eqnarray}
\beta_i^{\mu\nu}=\left(\begin{matrix}a_i&b_i&b_i&b_i\\b_i&a_i&b_i&b_i\\
b_i&b_i&a_i&b_i\\b_i&b_i&b_i&a_i\end{matrix}\right),\quad
\beta_j^{\mu\nu}=\left(\begin{matrix}a_j&b_j&b_j&b_j\\b_j&a_j&b_j&b_j\\
b_j&b_j&a_j&b_j\\b_j&b_j&b_j&a_j\end{matrix}\right).
\label{definition_beta}
\end{eqnarray}
Remember that the entries $a_i$,$a_j$,$b_i$ and $b_j$ of the matrix representation of the
tensors $\beta_i^{\mu\nu}$ and $\beta_j^{\mu\nu}$ in ($\ref{definition_beta}$), have
already been defined in ($\ref{coefficients_inverse_alpha}$).
Since a complete set of eigenstates to all four components of the momentum operator in position space
can be found ($\ref{momentum_eigenstates_A}$),($\ref{eigenvalue_equation_momentum}$), although the
components of the momentum operator do not commute with each other in case of the position
representation ($\ref{commutators_position_representation}$), the theorem that such a set of eigenstates
with respect to two operators does exist exactly then, if these commutators commute with each other,
does not hold anymore in the presented quaternionic generalization of quantum mechanics.
This property of course arises directly from the noncommutativity of the units of the
imaginary directions, $i$, $j$, $k$, of the quaternionic number space.

\section{Consequences for Quantum Field Theory}

\subsection{Calculation of the Generalized Propagator}

In this section are considered the corresponding consequences of the generalized quantization postulate of
quantum mechanics ($\ref{Quaternionic_Quantization_QM}$), which has been introduced in the last section, for
quantum field theory. Since with respect to the derivation of the propagator a scalar field is considered,
the generalized quantization postulate has not to be transferred to field quantization here, but the influence
arising from the generalized quantum theoretical field equations to quantum field theory are explored. By
using the generalized expression of the plane waves as momentum eigenstates given in
($\ref{momentum_eigenstates_A}$) and ($\ref{momentum_eigenstates_B}$) respectively,
a free scalar field as solution of the generalized Klein-Gordon equation reads

\begin{eqnarray}
\phi(x)&=&\int \frac{d^3 p}{\sqrt{\left(2\pi\right)^3 2p_0}} \left\{
q\exp\left[-i\beta_i^{\mu\nu}p_\mu x_\nu
-j\beta_j^{\mu\nu} p_\mu x_\nu\right]
+q^{*}\exp\left[i\beta_i^{\mu\nu}p_\mu x_\nu
+j\beta_j^{\mu\nu} p_\mu x_\nu\right]\right\}.
\end{eqnarray}
The quaternionic quantization principle of quantum mechanics ($\ref{Quaternionic_Quantization_QM}$) influences
the shape of free quantum theoretical field equations and thus the plane waves, but it does not influence the
Fock space structure of the Hilbert space of many particles. Therefore the postulated commutation relations
between the coefficients of the plane waves concerning field quantization remain the same. This means that
to obtain the quantum properties of a scalar field in case of quaternionic quantization, the coefficients
have to become operators, $q \rightarrow \hat q, q^{*} \rightarrow \hat q^{\dagger}$, which have the same
properties as in the usual case,

\begin{equation}
\left[\hat q\left(p\right),\hat q^{\dagger}\left(p^{\prime}\right)\right]
=\delta\left(p-p^{\prime}\right).
\end{equation}
The corresponding scalar field operator reads

\begin{eqnarray}
\hat \phi(x)&=&\int \frac{d^3 p}{\sqrt{\left(2\pi\right)^3 2p_0}}
\left\{\hat q \exp\left[-i\beta_i^{\mu\nu}p_\mu x_\nu-j\beta_j^{\mu\nu} p_\mu x_\nu\right]
+\hat q^{\dagger}\exp\left[i\beta_i^{\mu\nu}p_\mu x_\nu+j\beta_j^{\mu\nu} p_\mu x_\nu\right]
\right\}.
\label{quaternionic_quantum_field}
\end{eqnarray}
To obtain the corresponding propagator to this generalized quantum field ($\ref{quaternionic_quantum_field}$),
as usual one has to consider the expectation value with respect to the vacuum state, $|0\rangle$,
of the time ordered product,

\begin{equation}
T\left[A\left(t_1\right)B\left(t_2\right)\right]=
\begin{cases}
A\left(t_1\right)B\left(t_2\right),\quad {\rm if}\ t_1 > t_2\\
B\left(t_2\right)A\left(t_1\right),\quad {\rm if}\ t_2 > t_1
\end{cases},
\end{equation}
of the field operator at two different space-time points $x$ and $y$,

\begin{eqnarray}
&&G\left(x-y\right)=\langle 0|T\left[\hat \phi(x)\hat \phi(y)\right]|0 \rangle\nonumber\\
&&=\int \frac{d^3 p}{\left(2\pi\right)^3 2 p_0}
\left\{\theta\left(x_0-y_0\right)\exp\left[\mathcal{Z}\left(-\kappa\left(p,x\right),\kappa\left(p,y\right)\right) \right]
+\theta\left(y_0-x_0\right)\exp\left[\mathcal{Z}\left(-\kappa\left(p,y\right),\kappa(p,x)\right)
\right]\right\}\nonumber\\
&&=i\int\frac{d\tau}{2\pi}\int \frac{d^3 p}{\left(2\pi\right)^3}
\left\{\frac{\exp\left[-i\tau\left(x_0-y_0\right)\right]
\exp\left[\mathcal{Z}\left(-\kappa\left(p,x\right),\kappa\left(p,y\right)\right)\right]}{2p_0\tau+i\epsilon}
+\frac{\exp\left[-i\tau\left(y_0-x_0\right)\right]
\exp\left[\mathcal{Z}\left(-\kappa\left(p,y\right),\kappa\left(p,x\right)\right)
\right]}{2p_0\tau+i\epsilon}\right\}\nonumber\\
&&=i\int\frac{d\tau}{2\pi}\int \frac{d^3 p}{\left(2\pi\right)^3}
\frac{\exp\left[-i\tau\left(x_0-y_0\right)\right]}{2p_0\tau+i\epsilon}
\left\{\exp\left[\mathcal{Z}\left(-\kappa\left(p,x\right),\kappa\left(p,y\right)\right)\right]
-\exp\left[\mathcal{Z}\left(-\kappa\left(p,y\right),\kappa\left(p,x\right)\right)\right]\right\},
\label{quaternionic_propagator}
\end{eqnarray}
where has been defined

\begin{equation}
\kappa(p,x)=i\beta_i^{\mu\nu}p_\mu x_\nu+j\beta_j^{\mu\nu} p_\mu x_\nu.
\end{equation}
In ($\ref{quaternionic_propagator}$) have been used as usual the definition of the $\Theta$-function,

\begin{equation}
\Theta \left(x_0-y_0\right)=\lim_{\epsilon \to 0}\ -\frac{1}{2\pi i}\int d \tau \frac{\exp\left[
-i\tau\left(x_0-y_0\right)\right]}{\tau+i\epsilon},
\end{equation}
and the Baker-Campbell-Hausdorff formula,

\begin{equation}
\exp{A}\exp{B}=\exp{\mathcal{Z}\left(A,B\right)},\quad{\rm with}\quad
\mathcal{Z}\left(A,B\right)=\sum_{n=1}^{\infty}\frac{\left(-1\right)^{n-1}}{n}\sum_{i=1}^{n}
\sum_{r_i+s_i > 0}\prod_{m=1}^{n} \frac{A^{r_m}B^{s_m}}{r_m ! s_m !}.
\end{equation}

\subsection{Quaternionic Gauge Principle}

The wave function in the generalized Dirac equation ($\ref{quaternionic_Dirac_equation}$) represents a quaternionic
spinor wave function. This means that the spinor structure is the same as in usual quantum field theory, but the
wave function is quaternionic and thus it is of a shape as defined in ($\ref{quaternionic_wave_function}$). With
respect to usual complex wave functions one can perform phase transformation, $\psi_\mathbb{C}(x)\rightarrow
e^{i\alpha}\psi_\mathbb{C}(x)$. The free wave equations are invariant under such a transformation, if it is chosen
to be a global transformation. The postulate of local invariance under a phase transformation leads to the necessity
to introduce the electromagnetic potential and thus the electromagnetic interaction has its origin in a symmetry
principle. This symmetry principle has to be generalized in the quaternionic case, since in this case a transformation
is possible, which refers to all quaternionic directions, $i$, $j$ and $k$. The intricacy of such a generalization
consists in the noncommutativity between the quantities $i$, $j$ and $k$, which build a Lie Algebra, the Lie Algebra
belonging to the $SU(2)$ namely. Accordingly even usual electrodynamics has to be generalized to a certain kind of
non Abelian gauge theory. The quaternionic Dirac equation ($\ref{quaternionic_Dirac_equation}$) contains the
following generalized quaternionic phase invariance:

\begin{equation}
\psi\longrightarrow\exp(i\varphi+\varkappa j\chi+\varkappa k\rho)\psi,\quad
\alpha^{\mu\nu}\longrightarrow\exp(i\varphi+\varkappa j\chi+\varkappa k\rho)
\alpha^{\mu\nu}\exp(-i\varphi-\varkappa j\chi-\varkappa k\rho).
\label{quaternionic_phase_transformations}
\end{equation}
The prefactor $\varkappa$ of the quantities $j$ and $k$ maintains that the theory becomes approximatively
equal to usual electrodynamics, if $\varkappa$ goes to zero.    
Since $\alpha^{\mu\nu}$ is a quaternionic tensor, it has of course to be transformed as well.
A precondition for the invariance of the quaternionic Dirac equation ($\ref{quaternionic_Dirac_equation}$)
under the generalized quaternionic phase transformations ($\ref{quaternionic_phase_transformations}$)
is that the $\gamma$-matrices commute with the quaternionic quantization tensor $\alpha^{\mu\nu}$
($\ref{commutator_alpha_gamma}$).
If this symmetry is postulated to be a local symmetry, then the corresponding potential as generalization
of the usual electromagnetic potential has to contain three components referring to the three
directions of the quaternionic space. This means that the Lagrangian containing the local quaternionic phase
invariance reads as follows:

\begin{equation}
\mathcal{L}=\bar \psi \left(\alpha^{\mu\nu}\gamma_\mu \mathcal{D}_\nu-m\right)\psi,
\label{local_Dirac_Lagrangian}
\end{equation}
where the covariant derivative $\mathcal{D}_\mu$ is defined in the following way:

\begin{equation}
\mathcal{D}_\mu=\partial_\mu+i\mathcal{A}_\mu+\varkappa j\mathcal{B}_\mu+\varkappa k\mathcal{C}_\mu.
\label{covariant_derivative}
\end{equation}
The gauge potentials appearing in ($\ref{covariant_derivative}$) under an infinitesimal
transformation have to transform according to

\begin{eqnarray}
&&\mathcal{A}_\mu \longrightarrow \mathcal{A}_\mu-\partial_\mu \varphi
-\varkappa k \mathcal{A}_\mu \chi+\varkappa j \mathcal{A}_\mu \rho,
\nonumber\\&&
\mathcal{B}_\mu \longrightarrow \mathcal{B}_\mu-\partial_\mu \chi
-\varkappa i \mathcal{B}_\mu \rho+k \mathcal{B}_\mu \varphi,
\nonumber\\&&
\mathcal{C}_\mu \longrightarrow \mathcal{C}_\mu-\partial_\mu \rho
-j \mathcal{C}_\mu \varphi+\varkappa i \mathcal{C}_\mu \chi,
\label{transformation_quaternionic_potential}
\end{eqnarray}
if a local phase transformation is considered. This means that the Lagrangian ($\ref{local_Dirac_Lagrangian}$)
is invariant under combined local transformations of the shape ($\ref{quaternionic_phase_transformations}$) and 
($\ref{transformation_quaternionic_potential}$) and represents the generalization of electromagnetism
with respect to the quaternionic generalization of quantum mechanics according to this paper.
Accordingly also a generalized field strength tensor has to be built based on the generalized covariant
derivative ($\ref{covariant_derivative}$), which is as usual defined as commutator of the components
of the covariant derivative and accordingly reads

\begin{eqnarray}
\mathcal{F}_{\mu\nu}=\left[\mathcal{D}_\mu,\mathcal{D}_\nu\right]
&=&i\left[\partial_\mu \mathcal{A}_\nu-\partial_\nu \mathcal{A}_\mu
+\varkappa^2\left(\mathcal{B}_\mu \mathcal{C}_\nu-\mathcal{C}_\mu \mathcal{B}_\nu\right)\right]
+\varkappa j\left[\partial_\mu \mathcal{B}_\nu-\partial_\nu \mathcal{B}_\mu
+\mathcal{C}_\mu \mathcal{A}_\nu-\mathcal{A}_\mu \mathcal{C}_\nu \right]
\nonumber\\
&&+\varkappa k\left[\partial_\mu \mathcal{C}_\nu-\partial_\nu \mathcal{C}_\mu
+\mathcal{A}_\mu \mathcal{B}_\nu-\mathcal{B}_\mu \mathcal{A}_\nu\right]\nonumber\\
&\equiv& i\mathcal{I}_{\mu\nu}+\varkappa j \mathcal{J}_{\mu\nu}
+\varkappa k \mathcal{K}_{\mu\nu},
\label{generalized_field_strength}
\end{eqnarray}
where the last line ($\ref{generalized_field_strength}$) serves as a definition of the several components
$\mathcal{I}_{\mu\nu}$, $\mathcal{J}_{\mu\nu}$ and $\mathcal{K}_{\mu\nu}$. In analogy to the usual case
it is suggesting to postulate the following Lagrangian for the interaction fields of
the generalized electrodynamics,

\begin{equation}
\mathcal{L}=\frac{1}{4}\mathcal{F}^{*}_{\mu\nu}\mathcal{F}^{\mu\nu}
=\frac{1}{4}\mathcal{I}_{\mu\nu}\mathcal{I}^{\mu\nu}
+\frac{1}{4}\varkappa^2 \mathcal{J}_{\mu\nu}\mathcal{J}^{\mu\nu}
+\frac{1}{4}\varkappa^2 \mathcal{K}_{\mu\nu}\mathcal{K}^{\mu\nu}.
\end{equation}

\section{Canonical Quantization of General Relativity with Quaternions}

\subsection{Quaternionic Quantization in Quantum Geometrodynamics}

In the last two sections the quaternionic quantization principle has been considered with respect to quantum mechanics
and the corresponding consequences for quantum field theory. But the main interest of this extension of quantum theory
arises, if it is explored within the quantum description of general relativity. Since the appropriate quantum description
of general relativity has not been found yet, it could indeed be possible that the generalization of the concept of quantization, which is suggested in this paper, is necessary to incorporate also general relativity to a quantum description
of all interactions, although the approximation with $\varkappa \rightarrow 0$ was appropriate to treat the physics
of elementary particles at low energies, which is based on the other fundamental interactions in nature.
Of course, the generalization of the quantization concept has been formulated with respect to canonical quantization.
Accordingly, a modification of the canonical quantization of general relativity is considered in this section and in
the next section the corresponding modification is extended to canonical quantum supergravity. The canonical quantization
of general relativity is based on a foliation of space-time into a spacelike three dimensional submanifold $\Sigma$
and one separated space-time direction described by $\tau$, which is considered as time-coordinate. Accordingly the
metric $g_{\mu\nu}$ referring to the complete space-time can be splitted into a part referring to the spacelike
submanifold $\Sigma$, which is denoted as $h_{ab}$, and the other components, which are related to the time coordinate
$\tau$ and are expressed by the lapse function $N$ and the shift vector $N_a$, see \cite{Kiefer:2004gr} for example.
The complete metric expressed by these variables reads as follows:

\begin{equation}
g_{\mu\nu}=\left(\begin{matrix}N_a N^a-N^2 & N_b\\ N_c & h_{ab}\end{matrix}\right).
\label{splitting_metric}
\end{equation}
If the representation ($\ref{splitting_metric}$) of the metric is used, then the Einstein-Hilbert action
can be written in the following way:

\begin{eqnarray}
S_{EH}&=&\int_{\mathcal{M}}dt\ d^3 x\ \mathcal{L}_g=\frac{1}{16 \pi G}\int_{\mathcal{M}}dt\ d^3 x\ N
\left(G^{abcd}K_{ab}K_{cd}+\sqrt{h}\left[R_h-2\Lambda\right]\right),
\label{Einstein-Hilbert_A}
\end{eqnarray}
where $G$ denotes the gravitational constant, $R_h$ denotes the part of the Ricci scalar built from the three metric
$h_{ab}$ and thus refers to the submanifold $\Sigma$ and $K_{ab}$ denotes the extrinsic curvature,
which is defined as

\begin{equation}
K_{ab}=\frac{1}{2N}\left(\dot h_{ab}-D_a N_b-D_b N_a\right),
\label{extrinsic_curvature}
\end{equation}
and the forth order tensor $G_{abcd}$ is called DeWitt metric and is defined as 

\begin{equation}
G_{abcd}=\frac{1}{2\sqrt{h}}\left(h_{ac}h_{bd}+h_{ad}h_{bc}-h_{ab}h_{cd}\right).
\end{equation}
By referring to the Lagrangian $\mathcal{L}_g$ within ($\ref{Einstein-Hilbert_A}$), the canonical conjugated
quantity $\pi^{ab}$ can be defined,

\begin{equation}
\pi^{ab}=\frac{\partial \mathcal{L}_g}{\partial \dot h_{ab}}=\frac{\sqrt{h}}{16 \pi G}\left(K^{ab}-K h^{ab}\right),
\label{canonical_conjugated_momentum}
\end{equation}
and by using the canonical conjugated momentum ($\ref{canonical_conjugated_momentum}$) the Einstein-Hilbert action
($\ref{Einstein-Hilbert_A}$) can be reexpressed to

\begin{equation}
S_{EH}=\frac{1}{16 \pi G}\int_{\mathcal{M}}dt\ d^3 x \left(\pi^{ab}\dot h_{ab}-N\mathcal{H}_\tau-N^a \mathcal{H}_a\right),
\label{Einstein-Hilbert_B}
\end{equation}
where $\mathcal{H}_\tau$ denotes the part of the Hamiltonian density referring to the time direction $\tau$ and
$\mathcal{H}_a$ denotes the part of the Hamiltonian density referring to the submanifold $\Sigma$. 
Variation of ($\ref{Einstein-Hilbert_B}$) with respect to the three metric yields the dynamical constraints,
the Hamiltonian constraint and the diffeomorphism constraint,

\begin{eqnarray}
\mathcal{H}_\tau=16 \pi G G_{abcd} \pi^{ab}\pi^{cd}-\frac{\sqrt{h}}{16 \pi G}\left(R_h-2 \Lambda\right)=0,\quad
\mathcal{H}_a=-2 D_b  \pi_{\ a}^b=0.
\label{constraints}
\end{eqnarray}
To obtain a quantum description of canonical general relativity, the three metric $h_{ab}$ defined in
($\ref{splitting_metric}$) as well as the corresponding canonical conjugated variable $\pi_{ab}$
defined in ($\ref{canonical_conjugated_momentum}$) have to be converted to operators,
$h_{ab}\rightarrow \hat h_{ab}$ and $\pi^{ab}\rightarrow \hat \pi_{ab}$. Within the usual description of
quantum geometrodynamics one postulates in analogy to the usual Heisenbergian commutation relation between
position and momentum as quantization principle in quantum mechanics the following commutation relations
between the operator describing the three metric $\hat h_{ab}$ and the operator describing the
corresponding canonical conjugated momentum $\hat \pi^{ab}$,

\begin{equation}
\left[\hat h_{ab}(x),\hat \pi^{cd}(y)\right]=\frac{i}{2}\left(\delta^c_a \delta^d_b+\delta^c_b \delta^d_a\right)\delta(x-y).
\label{usual_quantization_GR}
\end{equation}
If now the assumption is made that general quantum theory contains a quaternionic quantization principle,
then this quantization principle becomes not only manifest with respect to quantum mechanics, but also
a quantum description of general relativity has to be based on a corresponding quantization. This means
that the quaternionic quantization principle ($\ref{Quaternionic_Quantization_QM}$) as it has been formulated
as generalization of quantum mechanics has to be transferred to the variables of canonical quantum gravity
and thus ($\ref{usual_quantization_GR}$) has to be generalized. To be analogue to the case of quantum
mechanics, the generalization has to be performed in such a way that the commutation relations between
the components of the three metric $\hat h_{ab}$ and the corresponding components of the canonical conjugated
variable $\hat \pi^{ab}$ remain the same and the commutation relations between the components of $\hat h_{ab}$
and all the other components of $\hat \pi^{ab}$ variable are equal to $j$ times the parameter $\varkappa$.
This leads to the following transition of the commutation relation between
$\hat h_{ab}$ and $\hat \pi^{ab}$:

\begin{equation}
\left[\hat h_{ab}(x),\hat \pi^{cd}(y)\right]=\frac{i}{2}\left(\delta^c_a \delta^d_b+\delta^c_b \delta^d_a\right)\delta(x-y)
\quad\longrightarrow\quad\left[\hat h_{ab}(x),\hat \pi^{cd}(y)\right]=M_{ab}^{cd}\delta\left(x-y\right),
\label{Quaternionic_Quantization_GR}
\end{equation}
where has been introduced the new quaternionic tensor $M_{ab}^{cd}$ of fourth order in three dimensions,
which can be written as a three cross three matrix, which contains second order tensors as entries of
this matrix,

\begin{equation}
M_{ab}^{cd}=\frac{1}{2}\left(
\begin{matrix}
m_{a1}^{c1} & m_{a1}^{c2} & m_{a1}^{c3}\\
m_{a2}^{c1} & m_{a2}^{c2} & m_{a2}^{c3}\\
m_{a3}^{c1} & m_{a3}^{c2} & m_{a3}^{c3}\\
\end{matrix}\right).
\label{quaternionic_tensor_GR_A}
\end{equation}
The entries of ($\ref{quaternionic_tensor_GR_A}$) can again be written as three cross three
matrices, which contain quaternionic expressions and are of the following form:

\begin{eqnarray}
&&m_{a1}^{c1}=\left(\begin{matrix}
2i & 2\varkappa j & 2\varkappa j\\2\varkappa j & i & \varkappa j\\2\varkappa j & \varkappa j & i
\end{matrix}\right),\quad
m_{a1}^{c2}=\left(\begin{matrix}
2\varkappa j & 2\varkappa j & 2\varkappa j\\i & 2\varkappa j & \varkappa j\\\varkappa j & 2\varkappa j & \varkappa j
\end{matrix}\right),\quad
m_{a1}^{c3}=\left(\begin{matrix}
2\varkappa j & 2\varkappa j & 2\varkappa j\\\varkappa j & \varkappa j & 2\varkappa j\\i & \varkappa j & 2\varkappa j
\end{matrix}\right),\nonumber\\
&&m_{a2}^{c1}=\left(\begin{matrix}
2\varkappa j & i & \varkappa j\\2\varkappa j & 2\varkappa j & 2\varkappa j\\2\varkappa j & \varkappa j & \varkappa j
\end{matrix}\right),\quad
m_{a2}^{c2}=\left(\begin{matrix}
i & 2\varkappa j & \varkappa j\\2\varkappa j & 2i & 2\varkappa j\\\varkappa j & 2\varkappa j & i
\end{matrix}\right),\quad
m_{a2}^{c3}=\left(\begin{matrix}
\varkappa j & \varkappa j & 2\varkappa j\\2\varkappa j & 2\varkappa j & 2 \varkappa j\\\varkappa j & i & 2\varkappa j
\end{matrix}\right),\nonumber\\
&&m_{a3}^{c1}=\left(\begin{matrix}
2\varkappa j & \varkappa j & i\\2\varkappa j & \varkappa j & \varkappa j\\2\varkappa j & 2\varkappa j & 2\varkappa j
\end{matrix}\right),\quad
m_{a3}^{c2}=\left(\begin{matrix}
\varkappa j & 2\varkappa j & \varkappa j\\\varkappa j & 2\varkappa j & i\\2\varkappa j & 2\varkappa j & 2\varkappa j
\end{matrix}\right),\quad
m_{a3}^{c3}=\left(\begin{matrix}
i & \varkappa j & 2\varkappa j\\\varkappa j & i & 2\varkappa j\\2\varkappa j & 2\varkappa j & 2i
\end{matrix}\right).
\label{quaternionic_tensor_GR_B}
\end{eqnarray}
As in the usual case the factor $\frac{1}{2}$ of some entries arises from the symmetry property of the
three metric, $h_{ab}=h_{ba}$ leading also to $\pi^{ab}=\pi^{ba}$, implying that the components of the
components with different indices would appear doubly, if the corresponding factor two would not be
removed. As usual the operators act on states $|\Psi\rangle$, which are functionals depending on
$h_{ab}$ or $\pi^{ab}$ respectively, but take in analogy to the generalized states in quantum mechanics
quaternionic values. If the three metric representation of the operators defined
by the quaternionic quantization principle of general relativity is used
($\ref{Quaternionic_Quantization_GR}$), these operators read

\begin{equation}
\hat h_{ab}\left(x\right)|\Psi[h\left(x\right)]\rangle=h_{ab}\left(x\right)|\Psi[h\left(x\right)]\rangle,\quad
\hat \pi^{ab}\left(x\right)|\Psi[h\left(x\right)]\rangle=-M_{cd}^{ab}\frac{\delta}{\delta h_{cd}
\left(x\right)}|\Psi[h\left(x\right)]\rangle.
\end{equation}
The quaternionic quantization principle implies also nontrivial commutation relations between the components
of the three metric operator $\hat h_{ab}$ and the components of the operator of the canonical conjugated
quantity $\hat \pi^{ab}$. The commutation relations between the components of the three metric operator $\hat h_{ab}$
can be calculated by using the representation with respect to the canonical conjugated quantity reading
as follows:

\begin{equation}
\hat h_{ab}\left(x\right)|\Psi\left[\pi\left(x\right)\right]\rangle=M_{ab}^{cd}\frac{\delta}{\delta \pi^{cd}\left(x\right)}
|\Psi\left[\pi\left(x\right)\right]\rangle,\quad
\hat \pi^{ab}\left(x\right)|\Psi\left[\pi\left(x\right)\right]\rangle
=\pi^{ab}\left(x\right)|\Psi\left[\pi\left(x\right)\right]\rangle.
\end{equation}
The commutation relations between the components of the three metric arise from the fact that the components 
of the quaternionic tensor of fourth order, $M_{cd}^{ab}$, defining the quaternionic quantization principle ($\ref{Quaternionic_Quantization_GR}$) and being defined in ($\ref{quaternionic_tensor_GR_A}$) and ($\ref{quaternionic_tensor_GR_B}$) do not commute with each other as the components of the quaternionic tensor
in quantum mechanics $\alpha^{\mu}_{\ \nu}$ defined in ($\ref{quaternionic_tensor_QM}$).
Therefore to perform the calculation, the commutation relations between the components of the quaternionic
tensor of fourth order, $M_{cd}^{ab}$, are important. The components of $M_{cd}^{ab}$ fulfil the following
commutation relations with each other, which are analogue to the commutation relations referring to
$\alpha^{\mu}_{\ \nu}$ ($\ref{commutator_quantization_tensor_QM}$):

\begin{equation}
\left[M^{ab}_{cd},M^{ef}_{gh}\right]=\mathcal{M}^{abef}_{cdgh},
\end{equation}
where $\mathcal{M}^{abef}_{cdgh}$ is a tensor of eight order, which can be written as a matrix containing
tensors of sixth order,

\begin{equation}
\mathcal{M}^{abef}_{cdgh}=\varkappa \left(\begin{matrix}
\mu_{c1gh}^{a1ef} & \mu_{c2gh}^{a1ef} & \mu_{c3gh}^{a1ef}\\
\mu_{c1gh}^{a2ef} & \mu_{c2gh}^{a2ef} & \mu_{c3gh}^{a2ef}\\
\mu_{c1gh}^{a3ef} & \mu_{c2gh}^{a3ef} & \mu_{c3gh}^{a3ef}\\
\end{matrix}\right),
\end{equation}
where the tensors appearing in the matrix can be written as matrices again,

\begin{eqnarray}
&&\mu_{a1ef}^{c1gh}=\left[m_{a1}^{c1}, m_{ef}^{gh}\right]=
\left(\begin{matrix}
2m_{i\ ef}^{\ \ gh}& 2m_{j\ ef}^{\ \ gh}& 2m_{j\ ef}^{\ \ gh}\\
2m_{j\ ef}^{\ \ gh}& m_{i\ ef}^{\ \ gh}& m_{j\ ef}^{\ \ gh}\\
2m_{j\ ef}^{\ \ gh}& m_{j\ ef}^{\ \ gh}& m_{i\ ef}^{\ \ gh}\\
\end{matrix}\right),
\quad
\mu_{a1ef}^{c2gh}=\left[m_{a1}^{c2}, m_{ef}^{gh}\right]=
\left(\begin{matrix}
2m_{j\ ef}^{\ \ gh}& 2m_{j\ ef}^{\ \ gh}& 2m_{j\ ef}^{\ \ gh}\\
m_{i\ ef}^{\ \ gh}& 2m_{j\ ef}^{\ \ gh}& m_{j\ ef}^{\ \ gh}\\
m_{j\ ef}^{\ \ gh}& 2m_{j\ ef}^{\ \ gh}& m_{j\ ef}^{\ \ gh}\\
\end{matrix}\right),
\nonumber\\&&
\mu_{a1ef}^{c3gh}=\left[m_{a1}^{c3}, m_{ef}^{gh}\right]=
\left(\begin{matrix}
2m_{j\ ef}^{\ \ gh}& 2m_{j\ ef}^{\ \ gh}& 2m_{j\ ef}^{\ \ gh}\\
m_{j\ ef}^{\ \ gh}& m_{j\ ef}^{\ \ h}& 2m_{j\ ef}^{\ \ gh}\\
m_{i\ ef}^{\ \ gh}& m_{j\ ef}^{\ \ f1}& 2m_{j\ ef}^{\ \ gh}\\
\end{matrix}\right),
\quad
\mu_{a2ef}^{c1gh}=\left[m_{a2}^{c1}, m_{ef}^{gh}\right]=
\left(\begin{matrix}
2m_{j\ ef}^{\ \ gh}& m_{i\ ef}^{\ \ gh}& m_{j\ ef}^{\ \ gh}\\
2m_{j\ ef}^{\ \ gh}& 2m_{j\ ef}^{\ \ gh}& 2m_{j\ ef}^{\ \ gh}\\
2m_{j\ ef}^{\ \ gh}& m_{j\ ef}^{\ \ gh}& m_{j\ ef}^{\ \ gh}\\
\end{matrix}\right),
\nonumber\\&&
\mu_{a2ef}^{c2gh}=\left[m_{a2}^{c2}, m_{ef}^{gh}\right]=
\left(\begin{matrix}
m_{i\ ef}^{\ \ gh}& 2m_{j\ ef}^{\ \ gh}& m_{j\ ef}^{\ \ gh}\\
2m_{j\ ef}^{\ \ gh}& 2m_{i\ ef}^{\ \ gh}& 2m_{j\ ef}^{\ \ gh}\\
m_{j\ ef}^{\ \ gh}& 2m_{j\ ef}^{\ \ gh}& m_{i\ ef}^{\ \ gh}\\
\end{matrix}\right),
\quad
\mu_{a2ef}^{c3gh}=\left[m_{a2}^{c3}, m_{ef}^{gh}\right]=
\left(\begin{matrix}
m_{j\ ef}^{\ \ gh}& m_{j\ ef}^{\ \ gh}& 2m_{j\ ef}^{\ \ gh}\\
2m_{j\ ef}^{\ \ gh}& 2m_{j\ ef}^{\ \ gh}& 2m_{j\ ef}^{\ \ gh}\\
m_{j\ ef}^{\ \ gh}& m_{i\ ef}^{\ \ gh}& 2m_{j\ ef}^{\ \ gh}\\
\end{matrix}\right),
\nonumber\\&&
\mu_{a3ef}^{c1gh}=\left[m_{a3}^{c1}, m_{ef}^{gh}\right]=
\left(\begin{matrix}
2m_{i\ ef}^{\ \ gh}& m_{j\ ef}^{\ \ gh}& m_{i\ ef}^{\ \ gh}\\
2m_{j\ ef}^{\ \ gh}& m_{j\ ef}^{\ \ gh}& m_{j\ ef}^{\ \ gh}\\
2m_{j\ ef}^{\ \ gh}& 2m_{j\ ef}^{\ \ gh}& 2m_{j\ ef}^{\ \ gh}\\
\end{matrix}\right),
\quad
\mu_{a3ef}^{c2gh}=\left[m_{a3}^{c2}, m_{ef}^{gh}\right]=
\left(\begin{matrix}
m_{j\ ef}^{\ \ gh}& 2m_{j\ ef}^{\ \ gh}& m_{j\ ef}^{\ \ gh}\\
m_{j\ ef}^{\ \ gh}& 2m_{j\ ef}^{\ \ gh}& m_{i\ ef}^{\ \ gh}\\
2m_{j\ ef}^{\ \ gh}& 2m_{j\ ef}^{\ \ gh}& 2m_{j\ ef}^{\ \ gh}\\
\end{matrix}\right),
\nonumber\\&&
\mu_{a3ef}^{c3gh}=\left[m_{a3}^{c3}, m_{ef}^{gh}\right]=
\left(\begin{matrix}
m_{i\ ef}^{\ \ gh}& m_{j\ ef}^{\ \ gh}& 2m_{j\ ef}^{\ \ gh}\\
m_{j\ ef}^{\ \ gh}& m_{i\ ef}^{\ \ gh}& 2m_{j\ ef}^{\ \ gh}\\
2m_{j\ ef}^{\ \ gh}& 2m_{j\ ef}^{\ \ gh}& 2m_{i\ ef}^{\ \ gh}\\
\end{matrix}\right),
\end{eqnarray}
which again contain tensors of fourth order, which entries are defined as 

\begin{eqnarray}
m_{i\ a1}^{\ \ c1}&=&\left[i, m_{a1}^{c1}\right]
=\left(\begin{matrix}0 & 4k & 4k\\4k & 0 & 2k\\4k & 2k & 0\end{matrix}\right),\quad
m_{i\ a1}^{\ \ c2}=\left[i, m_{a1}^{c2}\right]
=\left(\begin{matrix}4k & 4k & 4k\\0 & 4k & 2k\\2k & 4k & 2k\end{matrix}\right),\quad
m_{i\ a1}^{\ \ c3}=\left[i, m_{a1}^{c3}\right]
=\left(\begin{matrix}4k & 4k & 4k\\2k & 2k & 4k\\0 & 2k & 4k\end{matrix}\right),
\nonumber\\
m_{i\ a2}^{\ \ c1}&=&\left[i, m_{a2}^{c1}\right]
=\left(\begin{matrix}4k & 0 & 2k\\4k & 4k & 4k\\4k & 2k & 2k\end{matrix}\right),\quad
m_{i\ a2}^{\ \ c2}=\left[i, m_{a2}^{c2}\right]
=\left(\begin{matrix}0 & 4k & 2k\\4k & 0 & 4k\\2k & 4k & 0\end{matrix}\right),\quad
m_{i\ a2}^{\ \ c3}=\left[i, m_{a2}^{c3}\right]
=\left(\begin{matrix}2k & 2k & 4k\\4k & 4k & 4k\\2k & 0 & 4k\end{matrix}\right),
\nonumber\\
m_{i\ a3}^{\ \ c1}&=&\left[i, m_{a3}^{c1}\right]
=\left(\begin{matrix}4k & 2k & 0\\4k & 2k & 2k\\4k & 4k & 4k\end{matrix}\right),\quad
m_{i\ a3}^{\ \ c2}=\left[i, m_{a3}^{c2}\right]
=\left(\begin{matrix}2k & 4k & 2k\\2k & 4k & 0\\4k & 4k & 4k\end{matrix}\right),\quad
m_{i\ a3}^{\ \ c3}=\left[i, m_{a3}^{c3}\right]
=\left(\begin{matrix}0 & 2k & 4k\\2k & 0 & 4k\\4k & 4k & 0\end{matrix}\right),
\nonumber\\
m_{j\ a1}^{\ \ c1}&=&\left[j, m_{a1}^{c1}\right]
=\left(\begin{matrix}-4k & 0 & 0\\0 & -2k & 0\\0 & 0 & -2k\end{matrix}\right),\quad
m_{j\ a1}^{\ \ c2}=\left[j, m_{a1}^{c2}\right]
=\left(\begin{matrix}0 & 0 & 0\\-2k & 0 & 0\\0 & 0 & 0\end{matrix}\right),\quad
m_{j\ a1}^{\ \ c3}=\left[j, m_{a1}^{c3}\right]
=\left(\begin{matrix}0 & 0 & 0\\0 & 0 & 0\\-2k & 0 & 0\end{matrix}\right),
\nonumber\\
m_{j\ a2}^{\ \ c1}&=&\left[j, m_{a2}^{c1}\right]
=\left(\begin{matrix}0 & -2k & 0\\0 & 0 & 0\\0 & 0 & 0\end{matrix}\right),\quad
m_{j\ a2}^{\ \ c2}=\left[j, m_{a2}^{c2}\right]
=\left(\begin{matrix}-2k & 0 & 0\\0 & -4k & 0\\0 & 0 & -2k\end{matrix}\right),\quad
m_{j\ a2}^{\ \ c3}=\left[j, m_{a2}^{c3}\right]
=\left(\begin{matrix}0 & 0 & 0\\0 & 0 & 0\\0 & -2k & 0\end{matrix}\right),
\nonumber\\
m_{j\ a3}^{\ \ c1}&=&\left[j, m_{a3}^{c1}\right]
=\left(\begin{matrix}0 & 0 & -2k\\0 & 0 & 0\\0 & 0 & 0\end{matrix}\right),\quad
m_{j\ a3}^{\ \ c2}=\left[j, m_{a3}^{c2}\right]
=\left(\begin{matrix}0 & 0 & 0\\0 & 0 & -2k\\0 & 0 & 0\end{matrix}\right),\quad
m_{j\ a3}^{\ \ c3}=\left[j, m_{a3}^{c3}\right]
=\left(\begin{matrix} -2k & 0 & 0\\0 & -2k & 0\\0 & 0 & -4k \end{matrix}\right).\nonumber\\
\end{eqnarray}
Accordingly the components of the three metric operator $\hat h_{ab}$ fulfil the following
commutation relations:

\begin{equation}
\left[\hat h_{ab}\left(x\right),\hat h_{cd}\left(y\right)\right]=
\left[-M_{ab}^{ef}\frac{\delta}{\delta \pi_{ef}(x)},-M_{cd}^{gh}\frac{\delta}{\delta \pi_{gh}(y)}\right]
=\mathcal{M}^{efgh}_{abcd}\left(M^{-1}\right)_{ef}^{ij}\left(M^{-1}\right)_{gh}^{kl}
\hat h_{ij}\left(x\right)\hat h_{kl}\left(y\right).
\label{commutator_h}
\end{equation} 
The commutation relations between the components of the operator of the canonical conjugated quantity
$\hat \pi^{ab}$, which can be calculated by referring to the three metric representation, read as follows

\begin{equation}
\left[\hat \pi_{ab}\left(x\right),\hat \pi_{cd}\left(y\right)\right]=
\left[M_{ab}^{ef}\frac{\delta}{\delta h_{ef}\left(x\right)},M_{cd}^{gh}\frac{\delta}{\delta h_{gh}\left(y\right)}\right]
=\mathcal{M}^{efgh}_{abcd}\left(M^{-1}\right)_{ef}^{ij}\left(M^{-1}\right)_{gh}^{kl}
\hat \pi_{ij}\left(x\right) \hat \pi_{kl}\left(y\right).
\label{commutator_pi}
\end{equation}
It is remarkable that according to ($\ref{commutator_h}$) and ($\ref{commutator_pi}$) the components of the operators  
$\hat h_{ab}$ and $\hat \pi^{ab}$ respectively fulfil even nontrivial commutation relations with the corresponding
components of the operators at other space points, what induces a kind of nonlocality. This nonlocality has its
origin in the property that the noncommutativity in ($\ref{commutator_h}$) and ($\ref{commutator_pi}$) arises from
the noncommutativity of the quaternionic components and this structure does not depend on the space point.
To obtain the quantum constraints restricting the states, which are physically possible, the canonical variables
$h_{ab}$ and $\pi^{ab}$ appearing in ($\ref{constraints}$) have to be replaced by the corresponding
operators $\hat h_{ab}$ and $\hat \pi^{ab}$. This yields the following quaternionic Wheeler-DeWitt equation
as quantum theoretical analogon to the Hamiltonian constraint as well the quantum theoretical version of the
diffeomorphism constraint, which read as follows:

\begin{eqnarray}
\left[16 \pi G G_{abcd} M_{ef}^{ab} M_{gh}^{cd}\frac{\delta}{\delta h_{ef}}\frac{\delta}{\delta h_{gh}}
-\frac{\sqrt{h}}{16 \pi G}\left(R_h-2 \Lambda\right)\right]|\Psi\left[h\right] \rangle=0,\quad
2 D_b h_{ac} M_{de}^{bc}\frac{\delta}{\delta h_{de}}|\Psi\left[h\right] \rangle=0.\nonumber\\
\label{quantum_constraints}
\end{eqnarray}
The constraints ($\ref{quantum_constraints}$) restrict the space of states, which are dynamically possible,
$|\Psi\left[h\left(x\right)\right]\rangle$, to a subspace $\mathcal{V}_{dyn}$ of the space of all states
$\mathcal{V}$: $\mathcal{V}_{dyn} \subset \mathcal{V}$.
The problems concerning the definition of an inner product in quantum  geometrodynamics remain the same
as in the usual case. But of course, the quaternionic quantization principle can analogously be transferred
to the new formulation of Hamiltonian general relativity given in \cite{Ashtekar:1986yd},\cite{Ashtekar:1987gu},
on which loop quantum gravity is based, which has been developed in \cite{Rovelli:1989za},\cite{Rovelli:1994ge}.
This is done in the next subsection.

\subsection{Quaternionic Quantization of Ashtekars Variables}

The Hamiltonian formulation of general relativity based on Ashtekars variables contains the connection of the
gravitational field on the submanifold $\Sigma$ as the decisive quantity, which is expressed as special spin
connection. The canonical conjugated quantity is the tetrad field on $\Sigma$ multiplied with the square root
of the three metric. Concretely, the Ashtekar variables are defined as follows:

\begin{equation}
A^i_a=\frac{\Gamma^i_a+\beta K^i_a}{G},\quad E^a_i=\sqrt{h} e^a_i,
\end{equation}
where $K^i_a$ denotes the extrinsic curvature defined in ($\ref{extrinsic_curvature}$) and $\Gamma^i_a$
is defined as follows: $\Gamma^i_a=-\frac{1}{2}\omega_{ajk}\epsilon^{ijk}$,
with $\omega_{ajk}$ describing the spin connection. $\beta$ denotes the Immirzi parameter. Since the connection $A^i_a$
and the canonical conjugated variable $E^a_i$ contain only one space-time index, the quaternionic quantization principle
can directly be transferred from quantum mechanics to general relativity by using the quaternionic quantization tensor $\alpha^{\mu}_{\ \nu}$ with respect to its spatial part referring to $\Sigma$, which shall be called $P^a_{\ b}$,

\begin{equation}
P^a_{\ b}=\left(\begin{matrix}i & \varkappa j & \varkappa j\\ \varkappa j & i & \varkappa j \\ \varkappa j & \varkappa j & i
\end{matrix}\right),
\label{quaternionic_quantization_tensor_P}
\end{equation}
and postulating the following generalization of the commutation relation:

\begin{equation}
\left[\hat A_a^i\left(x\right),\hat E^b_j\left(y\right)\right]=8\pi \beta i\delta^b_a \delta^i_j \delta\left(x-y\right)
\quad\longrightarrow\quad
\left[\hat A_a^i\left(x\right),\hat E^b_j\left(y\right)\right]=8\pi \beta P^b_{\ a} \delta^i_j \delta\left(x-y\right).
\label{quaternionic_quantization_Ashtekar}
\end{equation}
The quantization principle ($\ref{quaternionic_quantization_Ashtekar}$) leads to the following representation
of the operators with respect to the connection,

\begin{eqnarray}
\hat A_a^i\left(x\right)|\Psi\left[A\left(x\right)\right]\rangle=A_a^i\left(x\right)
|\Psi\left[A\left(x\right)\right]\rangle,\quad
\hat E_i^a\left(x\right)|\Psi\left[A\left(x\right)\right]\rangle=-8\pi \beta P^a_{\ b}
\frac{\delta}{\delta A^i_b\left(x\right)}
|\Psi\left[A\left(x\right)\right]\rangle.
\label{Ashtekar_operators_connection-representation}
\end{eqnarray}
The inverse of the three dimensional quaternionic quantization tensor $P^a_{\ b}$ reads:

\begin{equation}
\left(P^{-1}\right)^a_{\ b}=
\left(\begin{matrix}
u_i i+u_j j & v_i i+v_j j & v_i i+v_j j\\
v_i i+v_j j & u_i i+u_j j & v_i i+v_j j\\
v_i i+v_j j & v_i i+v_j j & u_i i+u_j j
\end{matrix}\right),
\label{quaternionic_quantization_tensor_P_inverse}
\end{equation}
where the coefficients of the quaternionic entries of $\left(P^{-1}\right)^{a}_{\ b}$
are of the following shape:

\begin{equation}
u_i=-\frac{3\varkappa^2+1}{4\varkappa^4+5\varkappa^2+1},\quad
u_j=\frac{2\varkappa^3}{4\varkappa^4+5\varkappa^2+1},\quad
v_i=\frac{\varkappa^2}{4\varkappa^4+5\varkappa^2+1},\quad
v_j=-\frac{2\varkappa^3+\varkappa}{4\varkappa^4+5\varkappa^2+1}.
\end{equation}
Of course, analogue to the case of $\alpha^{\mu}_{\ \nu}$ as well as $M^{ab}_{cd}$, also the components $P^a_{\ b}$
fulfil nontrivial commutation relations,

\begin{equation}
\left[P^{ab},P^{cd}\right]=\Xi^{abcd},
\end{equation}
where $\Xi_{abcd}$ is a tensor of fourth order, which can be written as a matrix,

\begin{equation}
\Xi^{abcd}=\varkappa
\left(\begin{matrix}\xi_i^{ab} & \xi_j^{ab} & \xi_j^{ab}\\
\xi_j^{ab} & \xi_i^{ab} & \xi_j^{ab}\\
\xi_j^{ab} & \xi_j^{ab} & \xi_i^{ab}
\end{matrix}\right),
\label{tensor_Xi}
\end{equation}
which contains tensors of second order, $\xi_i^{ab}$ and $\xi_j^{ab}$, which can be written
as matrices again,

\begin{equation}
\xi_i^{ab}=\left(\begin{matrix} 0 & 2k & 2k\\ 2k & 0 & 2k\\ 2k & 2k & 0 \end{matrix}\right)
,\quad \xi_j^{ab}=\left(\begin{matrix}-2k & 0 & 0\\ 0 & -2k & 0\\ 0 & 0 & -2k \end{matrix}\right).
\label{tensors_xi}
\end{equation}
This leads of course also in this formulation to nontrivial commutation relations of the components of the
operator of the connection with each other as well as between the components the operator of the conjugated
variable with each other, which are analogue to ($\ref{commutator_h}$) and ($\ref{commutator_pi}$) and
are given by

\begin{equation}
\left[\hat A_a^i\left(x\right),\hat A_b^j\left(y\right)\right]=\Xi_{acbd}
\left(P^{-1}\right)^{ce}\left(P^{-1}\right)^{df}
\hat A_e^i\left(x\right)\hat A_f^j\left(y\right),
\end{equation}
and 

\begin{equation}
\left[\hat E^a_i\left(x\right),\hat E^b_j\left(y\right)\right]=\Xi^{acbd}\left(P^{-1}\right)_{ce}
\left(P^{-1}\right)_{df}\hat E^e_i\left(x\right) \hat E^f_j\left(y\right).
\end{equation}
If the special choice $\beta=i$ is made, the Hamiltonian constraint is given by

\begin{equation}
\hat {\mathcal{H}}_\tau|\Psi\left[A\right] \rangle=\epsilon^{ijk}\hat F_{kab}\hat E^a_i \hat E^b_j|\Psi\left[A\right]\rangle
=16\pi^2 \beta^2 \epsilon^{ijk} F_{kab} P^a_{\ c} P^b_{\ d}
\frac{\delta}{\delta A_c^i}\frac{\delta}{\delta A_d^j}|\Psi\left[A\right]\rangle=0,
\end{equation}
and the diffeomorphism constraint is given by

\begin{equation}
\hat {\mathcal{H}}_a|\Psi\left[A\right]\rangle=\hat F^i_{ab}\hat E^b_i|\Psi\left[A\right]\rangle
=-8\pi \beta F^i_{ab} P^b_{\ c}\frac{\delta}{\delta A_c^i}|\Psi\left[A\right]\rangle=0,
\end{equation}
where the field strength tensor $F_{ab}^i$ is defined as

\begin{equation}
F_{ab}^i=2G \partial_a A_b^i+G^2 \epsilon^{i}_{jk} A_a^j E_b^k.
\end{equation}
In case of the Ashtekar formulation, the Gauss constraint additionally appears, which
in case of the postulation of the quaternionic quantization principle reads

\begin{eqnarray}
\mathcal{D}_a \hat E^a_i|\Psi\left[A\right]\rangle
=\mathcal{D}_a P^{a}_b \frac{\delta}{\delta A_b^i}|\Psi\left[A\right]\rangle=0,\quad {\rm with}\quad
\mathcal{D}_a E^a_i=\partial_a E_i^a+G\epsilon_{ijk}A_a^j E^{ka}.
\end{eqnarray}
As approach for the formulation of an inner product, $\langle \ \cdot\ |\ \cdot\ \rangle$, in analogy to the usual
case one can postulate,

\begin{equation}
\langle \Phi|\Psi\rangle=\int_{\mathcal{V}_{dyn}}\mathcal{D}\mu
\left[A\right]\ \Phi^{*}\left[A\right] \Psi\left[A\right],
\label{inner_product_Ashtekar}
\end{equation}
which constitutes a Hilbert space $\mathcal{H}_{\mathcal{G}}$ based on $\mathcal{V}_{dyn}$. $\Psi^{*}\left[A\right]$
still denotes the quaternionic conjugated quantity to $\Psi\left[A\right]$. The operators $\hat A_a^i$
and $\hat E^a_i$, which can be represented in the connection representation according to  ($\ref{Ashtekar_operators_connection-representation}$), are self-adjoint with respect to
the inner product ($\ref{inner_product_Ashtekar}$).

\section{Quaternionic Quantization of Supergravity}

\subsection{The Quaternionic Quantization Principle in case of the Appearance of Dirac Brackets}

The quaternionic quantization principle has already been considered for the special cases of quantum mechanics
and usual canonical general relativity in the last sections. In this section the quaternionic quantization procedure
shall be transferred to an extension of general relativity, $\mathcal{N}=1$ supergravity namely. Of course, the
canonical formulation of supergravity has to be considered. The canonical quantization of supergravity
has first been developed in \cite{D'Eath:1984sh},\cite{D'Eath:1984rf}. Further developments can for example be found in
\cite{D'Eath:1993up},\cite{D'Eath:1993jr},\cite{D'Eath:1996qv},\cite{Cheng:1996an},\cite{Cheng:1994sr},\cite{Nicolai:1990vb},\cite{Nicolai:1992xx},\cite{Matschull:1993hy},\cite{Melosch:1997wm},\cite{Sano:1992cx},\cite{Gambini:1995db},\cite{Moniz:1995bf},\cite{Cheng:1995mg},\cite{Foussats:1997vh}.
The quantization procedure becomes more intricate in the case of supergravity, since because of the appearance
of second class constraints, the quantization is not performed by postulating the commutation relations being
equal to $i$ times the corresponding classical Poisson-brackets, but the Poisson brackets have to be replaced
by Dirac brackets. This means that some commutators are proportional to more complicated tensors than just a
product of delta functions and nontrivial commutators between the components of the operators of the variables
with each other can already arise in the usual case. Thus the question arises how the quaternionic quantization
principle has to be implemented, if Dirac brackets appear. Accordingly the quaternionic quantization principle
has to be formulated in a more general form and it is necessary to find a general setting of the quaternionic
quantization principle as generalization of general quantum theory. In quantum mechanics the transition from
the usual quantization principle to the quaternionic quantization principle is performed by replacing the
Kronecker symbol $\delta^{\mu}_{\ \nu}$ by the quaternionic quantization tensor $\alpha^{\mu}_{\ \nu}$,
what has been done in ($\ref{Quaternionic_Quantization_QM}$). For an arbitrary vector $A_a$ and its
corresponding canonical conjugated quantity $B_b$, which are quantized by transferring Poisson brackets
to commutators, this means

\begin{equation}
\left[\hat A_a, \hat B_b \right]=i\delta_{ab}\quad \longrightarrow \quad \left[\hat A_a, \hat B_b \right]=P^D_{ab},
\label{transition_D}
\end{equation}
where $P^D_{ab}$ is the D-dimensional generalization of $P_{ab}$. It has now to be formulated the generalization of ($\ref{transition_D}$) to enable the transfer to theories, where second class constraints appear and thus Dirac
brackets have to be considered meaning that the commutator is postulated to be equal to a more complicated tensor
than the Kronecker symbol already in the usual case. To obtain such a formulation, it is helpful to reexpress
$P^D_{ab}$ as follows:

\begin{equation}
P^D_{ab}=i\mathcal{Q}_{ab}^{cd}\delta_{cd},
\end{equation}
where $\mathcal{Q}_{ab}^{cd}$ is a quaternionic tensor of fourth order, which reads

\begin{equation}
\mathcal{Q}_{ab}^{cd}=
\left(\begin{matrix}1 & -\varkappa k & ... & ...& -\varkappa k\\-\varkappa k & 1 & ...& ... & ...\\
... &... & ... & ... & ...\\ ... & ... & ... & 1 & -\varkappa k\\
-\varkappa k & ... & ... & -\varkappa k & 1 \end{matrix}\right)
\otimes
\frac{1}{D}
\left(\begin{matrix} 
1 & i & ... & ... & i\\ i & 1 & ... & ... & ...\\ ... & ... & ... & ... & ...\\
... & ... & ... & 1 & i\\ i & ... & ... & i & 1\\
\end{matrix}\right)
\equiv
\mathcal{P}_{ab} \otimes \frac{\mathcal{N}^{cd}}{D},\quad a,b,c,d=1...D,
\label{definition_Q}
\end{equation}
where $D$ denotes the number of dimensions of the space the quantities live in, which have to be quantized,
and $\mathcal{N}^{ab}$ has been chosen in such a way that $N^{ab}\delta_{ab}=D$, that it is invertible
and its non diagonal elements do not vanish anyhow, what is important with respect to the Dirac brackets.
In the special case $D=3$ it holds $i\mathcal{P}_{ab}=P_{ab}$. Remember that $P_{ab}$ has been defined in ($\ref{quaternionic_quantization_tensor_P}$). This means that the tensor $\mathcal{Q}_{ab}^{cd}$ serves as
a mediation tensor between the usual value of the commutator and the generalized value of the commutator.
Accordingly it is possible to generalize this quantization principle by transforming any tensor on the right
hand side of the usual quantization principle with the transformation tensor $\mathcal{Q}_{ab}^{cd}$.
This means for the quantization of an arbitrary vector $A_a$ and its corresponding canonical conjugated
quantity $B_b$, which are usually quantized by transferring the Dirac bracket to a commutator,

\begin{equation}
\left[\hat A_a,\hat B_b\right]_{+/-}=i\left\{A_a, B_b \right\}_D \quad \longrightarrow \quad
\left[\hat A_a,\hat B_b\right]_{+/-}=i\mathcal{Q}^{cd}_{ab}\left\{A_c, B_d \right\}_D,\quad
\label{transition}
\end{equation}
where $\left\{\ \cdot\ ,\ \cdot\ \right\}_D$ denotes the corresponding Dirac bracket being defined as

\begin{equation}
\left\{A,B\right\}_D=\left\{A,B\right\}_P
-\left\{A,\Theta_i\right\}_P \left(\Omega^{-1}\right)^{ij}\left\{\Theta_j,B\right\}_P,
\label{Dirac_bracket_definition}
\end{equation}
where $\left\{\ \cdot\ ,\ \cdot \ \right\}_P$ denotes the usual Poisson bracket, the $\Theta_i$
denote the second class constraints and $\Omega_{ij}=\left\{\Theta_i,\Theta_j\right\}_P$.
In case of $D=3$ as it appears with respect to the generalization of the quantization
of $\mathcal{N}=1$ supergravity in usual (3+1)-dimensional space-time, the quaternionic
mediation tensor $\mathcal{Q}_{ab}^{cd}$ multiplied with $i$, which has to be applied to the
value of the Dirac bracket to obtain the generalized commutator and is denoted by
$Q_{ab}^{cd}$, can be written as follows:

\begin{equation}
Q_{ab}^{cd}=i\mathcal{Q}_{ab}^{cd}=i\mathcal{P}_{ab}\otimes\frac{\mathcal{N}^{cd}}{3}
=P_{ab}\otimes \frac{\mathcal{N}^{cd}}{3}.
\end{equation}
Concerning the formulation of the generalized quantization rules arising from the Dirac
brackets, the inverse of $Q_{ab}^{cd}$ will have to be used, which can be written as

\begin{equation}
\left(Q^{-1}\right)^{ab}_{cd}=-\left(\mathcal{Q}^{-1}\right)^{ab}_{cd}i
=-3\left(\mathcal{P}^{-1}\right)^{ab}\otimes \left(\mathcal{N}^{-1}\right)_{cd}i
=3\left(P^{-1}\right)^{ab} \otimes \left(\mathcal{N}^{-1}\right)_{cd}.
\end{equation}
$\left(P^{-1}\right)^{ab}$ has been calculated in ($\ref{quaternionic_quantization_tensor_P_inverse}$).
The commutator of the several components of $Q_{ab}^{cd}$ with each other reads: 

\begin{equation}
\left[Q_{ab}^{cd},Q_{ef}^{gh}\right]=\Xi_{abef}\frac{\mathcal{N}^{cd}\mathcal{N}^{gh}}{9}\equiv \Gamma^{cdgh}_{abef}
+\frac{P_{ef}}{9}\Delta_{ab}^{gh}%\left[P_{ab},\mathcal{N}^{gh}\right]
\mathcal{N}^{cd}
-\frac{P_{ab}}{9}\Delta_{ef}^{cd}%\left[\mathcal{N}^{cd},P_{ef}\right]
\mathcal{N}^{gh},
\end{equation}
where the last equalization serves as a definition of $\Gamma^{cdgh}_{abef}$, $\Xi_{abcd}$
has already been determined in ($\ref{tensor_Xi}$) and $\Delta_{ab}^{cd}$ can be written
by using the definition ($\ref{tensors_xi}$) and thus reads as follows:
 
\begin{equation}
\Delta_{ab}^{cd}=\left(\begin{matrix}0&-\xi_{i\ ab}&-\xi_{i\ ab}
\\-\xi_{i\ ab}&0&-\xi_{i\ ab}
\\-\xi_{i\ ab}&-\xi_{i\ ab}&0\end{matrix}\right).
\end{equation}

\subsection{Quaternionic Quantization Principle in $\mathcal{N}=1$ Supergravity}

To formulate supergravity, the tetrad formalism has to be used. The tetrad field is related to the
metric field as usual by the following relation: $g_{\mu\nu}=\eta_{mn}e_\mu^m e_\nu^n$, where
$\eta_{mn}$ denotes the Minkowski metric. Concerning the canonical quantization as well as its
quaternionic extension, it is helpful to use the spinor representation of Minkowski vectors
given in \cite{Penrose:1985jw},\cite{Penrose:1986ca} and \cite{Stewart:1990uf} for example,
which is obtained by using the Pauli matrices. This means that the tetrad field $e_\mu^m$
can be rewritten as

\begin{equation}
e_\mu^{A A^{\prime}}=e_\mu^n \bar \sigma_n^{A A^{\prime}}.
\end{equation}
Here the matrices $\bar \sigma_n^{A A^{\prime}}$ are considered as components of a four-vector
containing the Pauli matrices and the negative unity matrix,

\begin{equation}
\bar \sigma_0=-\frac{1}{\sqrt{2}}\left(\begin{matrix}1 & 0\\ 0 & 1\end{matrix}\right),\quad
\bar \sigma^1=\frac{1}{\sqrt{2}}\left(\begin{matrix}0 & 1\\ 1 & 0\end{matrix}\right),\quad
\bar \sigma^2=\frac{1}{\sqrt{2}}\left(\begin{matrix}0 & -i\\ i & 0\end{matrix}\right),\quad
\bar \sigma^3=\frac{1}{\sqrt{2}}\left(\begin{matrix}1 & 0\\ 0 & -1\end{matrix}\right).
\end{equation}
Indices of the spinor representation are raised and lowered by the tensor $\epsilon^{AB}$,
which is the total antisymmetric tensor with respect to the spinor space.
The new quantity $n^{AA^{\prime}}$ directly related to the unit normal $n^\mu$ to the three
dimensional submanifold $\Sigma$ on which the three metric lives is defined as follows:

\begin{equation}
n^{AA^{\prime}}=e_\mu^{AA^{\prime}}n^\mu,
\end{equation}
and fulfils the following relations:

\begin{equation}
\quad n_{AA^{\prime}}e_a^{AA^{\prime}}=0,\quad n_{AA^{\prime}}n^{AA^{\prime}}=1.
\end{equation}
The unit normal, the components of the tetrad field and $N$ as well as $N^a$ are further related by

\begin{equation}
e_0^{AA^{\prime}}=N n^{AA^{\prime}}+N^a e_a^{AA^{\prime}},
\end{equation}
and the action of $\mathcal{N}=1$ supergravity is given by

\begin{eqnarray}
\mathcal{S}_{\mathcal{N}=1}&=&\frac{1}{16\pi G}\int d^4 x\ e\ e^\mu_a e^\nu_b R_{\mu\nu}^{ab}
+\frac{1}{2}\int d^4 x\ e\ \epsilon^{\mu\nu\rho\sigma}\left(\bar \psi_\mu^{A^{\prime}}
e_{A A^{\prime}\nu} D_\rho \psi_\sigma^{A}+h.c.\right),
\label{action_N=1_supergravity}
\end{eqnarray}
where $e=\det\left(e_\mu^m\right)$, $\epsilon^{\mu\nu\rho\sigma}$ denotes the total antisymmetric tensor in four
dimensions and the covariant derivative applied to a spinor reads

\begin{equation}
D_\mu \psi_\nu^{A}=\partial_\mu \psi_\nu^{A}+\omega_{\mu B}^{A}\psi_\nu^{B},
\end{equation}
where $\omega_{\mu B}^{A}$ denotes the spin connection in the spinor representation. Because of the appearance
of the Rarita Schwinger field the existence of a torsion is induced, 

\begin{equation}
\mathcal{T}^{AA^{\prime}}_{\mu\nu}=D_{[\mu} e_{\nu]}^{AA^{\prime}}=-4\pi i\bar \psi^{A^{\prime}}_{[\mu}\psi^A_{\nu ]},
\end{equation}
which is however assumed to vanish in the further consideration of this paper. The action of $\mathcal{N}=1$
supergravity ($\ref{action_N=1_supergravity}$) has three symmetries. Since it is a supersymmetric theory,
it of course contains a supersymmetry and is invariant under the following corresponding transformations:

\begin{equation}
\delta e_\mu^{A A^{\prime}}=-i\sqrt{8\pi G}
\left(\zeta^{A}\bar \psi_\mu^{A^{\prime}}+\bar \zeta^{A^{\prime}}\psi_\mu^{A}\right),\quad
\delta \psi_\mu^{A}=\frac{D_\mu \zeta^{A}}{\sqrt{2\pi G}},\quad \delta \bar \psi_\mu^{A^{\prime}}
=\frac{D_\mu \bar \zeta^{A^{\prime}}}{\sqrt{2\pi G}},
\end{equation}
where $\zeta^{A}$ and $\zeta^{A^{\prime}}$ denote the transformation parameter of the supersymmetry
and its adjoint. Besides it contains local Lorentz symmetry and accordingly is invariant under the
following transformations:

\begin{equation}
\delta e_\mu^{A A^{\prime}}=L_B^{\ A}e_\mu^{B A^{\prime}}+\bar L_{B^{\prime}}^{\ A^{\prime}}
e^{AB^{\prime}\mu},\quad \delta \psi_\mu^{A}=L_B^{\ A}\psi_\mu^B,\quad
\delta \bar \psi_\mu^{A^{\prime}}=\bar L_{B^{\prime}}^{\ A^{\prime}}\bar \psi_\mu^{B^{\prime}},
\end{equation}
where $L_B^{\ A}$ and $\bar L_B^{\ A^{\prime}}$ denote the transformation parameter of the
Lorentz group and its adjoint, and it contains local symmetry under coordinate transformations and
thus contains diffeomorphism invariance, which means that it is invariant under the following
transformations:

\begin{equation}
\delta e_\mu^{A A^{\prime}}=z^\nu \partial_\nu e_\mu^{A A^{\prime}}+e_\nu^{A A^{\prime}}\partial_\mu z^{\nu},\quad
\delta \psi_\mu^{A}=z^\nu \partial_\nu \psi_\mu^A+\psi_\nu^{A}\partial_\mu z^\nu,
\end{equation}
where $z^\mu$ denotes the corresponding local translation parameter.
The canonical conjugated variables to the tetrad field $e_\mu^{AA^{\prime}}$ as well as the Rarita Schwinger field
$\psi_a^{A}$ and its adjoint $\bar \psi_a^{A^{\prime}}$ are as usual defined with respect to the the action of
$\mathcal{N}=1$ supergravity ($\ref{action_N=1_supergravity}$) and accordingly the conjugated variable
to the tetrad field $p^a_{AA^{\prime}}$ is given by

\begin{equation}
p_{AA^{\prime}}^a=\frac{\delta \mathcal{S}_{\mathcal{N}=1}}{\delta \dot e_a^{AA^{\prime}}}.
\end{equation}
$p_{AA^{\prime}}^a$ is related to the canonical conjugated variable of $h_{ab}$ within the usual formulation
of canonical general relativity used in quantum geometrodynamics by

\begin{equation}
{\pi}^{ab}=\frac{e^{AA^{\prime}a}p_{AA^{\prime}}^b+e^{AA^{\prime}b}p_{AA^{\prime}}^a}{2},
\end{equation}
and the canonical conjugated variables to $\psi_a^{A}$ and $\psi_a^{A^{\prime}}$ read as follows:

\begin{equation}
\left(\pi_\psi\right)_A^a=\frac{\delta \mathcal{S}_{\mathcal{N}=1}}{\delta \dot \psi_a^A}=-\frac{1}{2}\epsilon^{abc}
\bar \psi_b^{A^{\prime}}e_{AA^{\prime}c},\quad
\left(\tilde \pi_{\bar \psi}\right)_{A^{\prime}}^a=\frac{\delta \mathcal{S}_{\mathcal{N}=1}}
{\delta \dot {\bar \psi}_a^{A^{\prime}}}
=\frac{1}{2}\epsilon^{abc}\psi_b^{A}e_{AA^{\prime}c}.
\end{equation}
The complete Hamiltonian of $\mathcal{N}=1$ supergravity corresponding to the action
($\ref{action_N=1_supergravity}$) reads

\begin{equation}
H=\int d^3 x \left(N\mathcal{H}_\tau+N^a\mathcal{H}_a
+\psi_0^A S_A+\bar S_{A^{\prime}}\bar \psi_0^{A^{\prime}}
-\omega_{AB0}J^{AB}-\bar \omega_{A^{\prime}B^{\prime}0}\bar J^{A^{\prime}B^{\prime}}\right),
\label{Hamiltonian_N=1_supergravity}
\end{equation}
which corresponds to the following constraints:
        
\begin{equation}
J_{AB}=0,\quad \bar J_{A^{\prime}B^{\prime}}=0,\quad  \mathcal{H}=0,\quad \mathcal{H}_a=0,\quad
S_A=0,\quad \bar S_{A^{\prime}}=0,
\label{constraints}
\end{equation}
if it is varied with respect to $N$, $N^{a}$, $\psi_0^{A}$, $\bar \psi_0^{A^{\prime}}$, $\omega_{AB0}$
$\bar \omega_{A^{\prime}B^{\prime}0}$. The part of the Hamiltonian density $\mathcal{H}_\tau$ of
($\ref{Hamiltonian_N=1_supergravity}$) belonging to the time direction reads under the
condition that the torsion vanishes

\begin{equation}
\mathcal{H}_\tau=16 \pi G G_{abcd}{\pi}^{ab}{\pi}^{cd}-\frac{\sqrt{h}R}{16 \pi G}
+\left(\frac{1}{2}\epsilon^{abc}\bar \psi_a^{A^{\prime}}n_{AA^{\prime}}\mathcal{D}_b \psi_c^A+h.c.\right),
\end{equation}
and the part of the Hamiltonian density $\mathcal{H}_a$ belonging to the spacelike
directions with the submanifold $\Sigma$ then reads

\begin{equation}
\mathcal{H}_a=-2 h_{ab} \mathcal{D}_c {\pi}^{bc}+\left(\frac{1}{2}\epsilon^{bcd}\bar \psi_b^{A^{\prime}}e_{AA^{\prime} a}\mathcal{D}_c \psi_d^A+h.c.\right).
\end{equation}
The expression $S_A$ and its conjugated quantity $\bar S_A^{\prime}$ represent the generators of
supersymmetry transformations and are defined as

\begin{eqnarray}
S_A&=&\epsilon^{abc}e_{A A^{\prime} a}\mathcal{D}_b \bar \psi^{A^{\prime}}_c
-i4\pi G p_{A A^{\prime}}^a \bar \psi^{A^{\prime}}_a,\quad%\nonumber\\
\bar S_{A^{\prime}}=\epsilon^{abc} e_{A A^{\prime} a}\mathcal{D}_b \psi^{A}_c
+i4\pi G p_{A A^{\prime}}^{a} \psi^{A}_a.
\label{supersymmetry_generators}
\end{eqnarray}
The expression $J_{AB}$ and its conjugated quantity are the generators of Lorentz transformations
and are defined as

\begin{eqnarray}
J_{AB}&=&e_{(A}^{A^{\prime} a}p_{B) A^{\prime} a}+\psi_{(A}^a \left(\pi_\psi\right)_{B)a}
=e_{(A}^{A^{\prime} a}p_{B) A^{\prime} a}
-\frac{1}{2}\psi_{(A}^a \epsilon_{abc}\bar \psi^{A^{\prime}b} e_{B) A^{\prime}}^c,\nonumber\\
\bar J_{A^{\prime}B^{\prime}}&=&e_{(A^{\prime}}^{A a}p_{B^{\prime}) A a}
+\bar \psi_{(A^{\prime}}^a \left(\tilde \pi_{\bar \psi}\right)_{B^{\prime})a}
=e_{(A^{\prime}}^{A a}p_{B^{\prime}) A a}
+\frac{1}{2}\bar \psi_{(A^{\prime}}^a \epsilon_{abc}\psi^{A b} e_{B^{\prime}) A}^c.
\end{eqnarray}
As already mentioned at the beginning of the first subsection of this section, the appearance
of second class constraints in supergravity implies that one has to introduce Dirac brackets,
which replace the usual Poisson brackets concerning the formulation of the quantization
rules. The general definition of a Dirac bracket is given in ($\ref{Dirac_bracket_definition}$)
and in case of supergravity the Dirac brackets between the several quantities read

\begin{eqnarray}
\left\{e_a^{A A^{\prime}}\left(x\right),e_b^{B B^{\prime}}\left(y\right)\right\}_D&=&0,\nonumber\\
\left\{e_a^{A A^{\prime}}\left(x\right),p^b_{B B^{\prime}}\left(y\right)\right\}_D&=&
\epsilon^A_{\ B} \epsilon^{A^{\prime}}_{\ B^{\prime}}\delta^b_a \delta(x-y),\nonumber\\
\left\{p^a_{A A^{\prime}}\left(x\right),p^b_{B B^{\prime}}\left(y\right)\right\}_D&=&
\frac{1}{4}\epsilon^{bcd}\psi_{Bd}D_{AB^{\prime}ec}\epsilon^{aef}\bar \psi_{A^{\prime} f}\delta(x-y)+h.c.,\nonumber\\
\left\{\psi_a^A\left(x\right),\psi_b^B\left(y\right)\right\}_D&=&0,\nonumber\\
\left\{\psi_a^A\left(x\right),\bar \psi_b^{A^{\prime}}\left(y\right)\right\}_D&=&
-D_{ab}^{A A^{\prime}}\delta(x-y),\nonumber\\
\left\{e_a^{A A^{\prime}}\left(x\right),\psi_b^{B}\left(y\right)\right\}_D&=&0,\nonumber\\
\left\{p^a_{A A^{\prime}}\left(x\right),\psi_b^B\left(y\right)\right\}_D&=&
\frac{1}{2}\epsilon^{acd}\psi_{Ad}D^{B}_{A^{\prime} bc}\delta(x-y),
\label{Dirac_brackets_supergravity}
\end{eqnarray}
where the quantity $D_{ab}^{A A^{\prime}}$ has been defined according to

\begin{equation}
D_{ab}^{A A^{\prime}}=-\frac{2i}{\sqrt{h}}e_b^{A B^{\prime}}e_{B B^{\prime} a} n^{B A^{\prime}}.
\label{definition_D}
\end{equation}
The generators of supersymmetry ($\ref{supersymmetry_generators}$) fulfil the following Dirac brackets:

\begin{equation}
\left\{S_A\left(x\right),S_B\left(y\right)\right\}_D=0,\quad
\left\{\bar S_{A^{\prime}}\left(x\right),\bar S_{B^{\prime}}\left(y\right)\right\}_D=0,\quad 
\left\{S_A\left(x\right),\bar S_{A^{\prime}}\left(y\right)\right\}_D=i4\pi G\mathcal{H}_{AA^{\prime}}(x)\delta(x-y).
\end{equation}
In the usual quantization procedure, the Dirac brackets ($\ref{Dirac_brackets_supergravity}$) have to be converted to
commutators multiplied with $-i$. A quaternionic quantization is performed by referring to ($\ref{transition}$) and
accordingly the commutator has to be extended by applying the operator $\mathcal{Q}_{ab}^{cd}$ to it. But this holds
only for the commutation relations involving the components of the quantized variables and their corresponding
canonical conjugated quantities and not for the other commutation relations, which have to be derived from them
because of their dependence on them.
The choice, if commutation or anticommutation relations have to be used, is completely analogue to the
usual case and this means that commutation relations are assumed, if at least one of the variables is
Grassmann-even and otherwise anticommutation relations are assumed. This means that the following
commutation and anticommutation relations are postulated:

\begin{eqnarray}
\left[e_a^{A A^{\prime}}(x),p^b_{B B^{\prime}}(y)\right]_{-}&=&
i\delta_{ed}\delta^{bf}\mathcal{Q}^{ce}_{af}\epsilon^A_{\ B} \epsilon^{A^{\prime}}_{\ B^{\prime}}
\delta^d_c \delta(x-y),\quad%\nonumber\\
\left[\psi_a^A(x),\bar \psi_b^{A^{\prime}}(y)\right]_{+}=
-i\mathcal{Q}^{cd}_{ab}D_{cd}^{A A^{\prime}}\delta(x-y).
\label{generalized_quantization_supergravity_A}
\end{eqnarray}
Besides one is led to the following commutation relation, if one assumes the quantization 
principle ($\ref{transition}$) also to be valid with respect to the commutation relations
between the components of $p^a_{A A^{\prime}}(x)$ and $\psi_a^A(x)$ as commutation
relation between a field operator and the canonical conjugated field operator belonging
to another field operator,

\begin{eqnarray}
\left[p^a_{A A^{\prime}}(x),\psi_b^B(y)\right]_{-}&=&\frac{i}{2}\delta_{gf}\delta^{ae}\mathcal{Q}^{fh}_{eb}
\epsilon^{gcd}\psi_{Ad}D^{B}_{A^{\prime} hc}\delta(x-y).
\label{generalized_quantization_supergravity_AA}
\end{eqnarray}
Since the transition rule ($\ref{transition}$) to obtain the commutation relations of the corresponding quantum theory,
if the Dirac brackets of the classical theory are given, holds only with respect to the commutation relations between
the components of a field operator and the components of a canonical conjugated variable because the other commutation
relations depend on them, the Dirac brackets given in ($\ref{Dirac_brackets_supergravity}$) accordingly have to be
converted to commutation relations by deriving them from the properties of the operators $\hat e_a^{AA^{\prime}}$,
$\hat p^a_{AA^{\prime}}$, $\hat \psi_a^{AA^{\prime}}$  and $\hat {\bar \psi}^a_{AA^{\prime}}$. These operators
are already defined by the commutation relations ($\ref{generalized_quantization_supergravity_A}$) and ($\ref{generalized_quantization_supergravity_AA}$).
The operators $\hat e_a^{AA^{\prime}}$, $\hat p^a_{AA^{\prime}}$, $\hat \psi_a^{AA^{\prime}}$ and
$\hat {\bar \psi}^a_{AA^{\prime}}$ fulfilling the commutation relations ($\ref{generalized_quantization_supergravity_A}$) and ($\ref{generalized_quantization_supergravity_AA}$) can be represented by referring on the corresponding
states depending on $e_a^{AA^{\prime}}$ and $\psi_a^{AA^{\prime}}$,
$|\Psi \rangle=|\Psi\left[e,\psi\right]\rangle$, and in this representation the
operators $\hat \psi_a^{AA^{\prime}}$ and $\hat {\bar \psi}^a_{AA^{\prime}}$ are of the following shape:

\begin{eqnarray}
\hat \psi_a^{A}\left(x\right)|\Psi\left[e\left(x\right),\psi\left(x\right)\right]\rangle
&=&\psi_a^{A}\left(x\right)|\Psi\left[e\left(x\right),\psi\left(x\right)\right]\rangle,\quad%\nonumber\\
\hat {\bar \psi}_a^{A^{\prime}}(x)|\Psi\left[e\left(x\right),\psi\left(x\right)\right]\rangle=
-i\mathcal{Q}^{cd}_{ba}D_{cd}^{AA^{\prime}}\frac{\delta}{\delta \psi_b^{A}\left(x\right)}
|\Psi\left[e\left(x\right),\psi\left(x\right)\right]\rangle,\nonumber\\
\label{operators_A1}
\end{eqnarray}
and the operators $\hat e_a^{AA^{\prime}}$ and $\hat p^a_{AA^{\prime}}$ look as follows:

\begin{eqnarray}
\hat e^{A A^{\prime}}_a\left(x\right)|\Psi\left[e\left(x\right),\psi\left(x\right)\right]\rangle
&=&e^{A A^{\prime}}_a\left(x\right)|\Psi\left[e\left(x\right),\psi\left(x\right)\right]\rangle,\\
\hat p_{A A^{\prime}}^a \left(x\right)|\Psi\left[e\left(x\right),\psi\left(x\right)\right]\rangle
&=&\left[-i\delta_{ed}\delta^{af}\mathcal{Q}^{ce}_{bf}\delta^{d}_c \frac{\delta}{\delta e_b^{A A^{\prime}}(x)}
-\frac{i}{2}\delta_{gf}\delta^{ae}\mathcal{Q}^{fh}_{eb}\epsilon^{gdc}\psi_{Ad}D^{B}_{A^{\prime} hc}
\frac{\delta}{\delta \psi^B_b\left(x\right)}\right]
|\Psi\left[e\left(x\right),\psi\left(x\right)\right]\rangle.\nonumber
\label{operators_A2}
\end{eqnarray}
If the states are assumed to depend on $\hat p^a_{AA^{\prime}}$ and $\hat {\bar \psi}^a_{AA^{\prime}}$,
$|\Psi\rangle=|\Psi\left[p\left(x\right),\bar \psi\left(x\right)\right]\rangle$,
then the operators $\hat \psi_a^{AA^{\prime}}$ and $\hat {\bar \psi}^a_{AA^{\prime}}$
can be represented as

\begin{eqnarray}
\hat \psi_a^{A}(x)|\Psi\left[p\left(x\right),\bar \psi\left(x\right)\right]\rangle
&=&-i\mathcal{Q}^{cd}_{ba}D_{cd}^{A A^{\prime}}\frac{\delta}{\delta \bar \psi_b^{A^{\prime}}\left(x\right)}
|\Psi\left[p\left(x\right),\bar \psi\left(x\right)\right]\rangle,\quad%\nonumber\\
\hat {\bar \psi}_a^{A}(x)|\Psi\left[p\left(x\right),\bar \psi\left(x\right)\right]\rangle
=\bar \psi_a^{A}(x)|\Psi\left[p\left(x\right),\bar \psi\left(x\right)\right]\rangle,\nonumber\\
\label{operators_B1}
\end{eqnarray}
and the operators $\hat e_a^{AA^{\prime}}$ and $\hat p^a_{AA^{\prime}}$ look as follows:

\begin{eqnarray}
\hat e_a^{A A^{\prime}}|\Psi\left[p\left(x\right),\bar \psi\left(x\right)\right]\rangle
&=&i\delta_{ed}\delta^{bf}\mathcal{Q}^{ce}_{af}\delta^{d}_c \frac{\delta}{\delta p^b_{A A^{\prime}}(x)}
|\Psi\left[p\left(x\right),\bar \psi\left(x\right)\right]\rangle,\nonumber\\
\hat p_{A A^{\prime}}^a \left(x\right)|\Psi\left[p\left(x\right),\bar \psi\left(x\right)\right]\rangle
&=&p_{A A^{\prime}}^a \left(x\right)|\Psi\left[p\left(x\right),\bar \psi\left(x\right)\right]\rangle.
\label{operators_B2}
\end{eqnarray}
The other commutation relations representing no independent assumptions can now be derived from the operators ($\ref{operators_A1}$), ($\ref{operators_A2}$), ($\ref{operators_B1}$) and ($\ref{operators_B2}$), which are
already defined through the commutation relations considered in ($\ref{generalized_quantization_supergravity_A}$)
and ($\ref{generalized_quantization_supergravity_AA}$). Accordingly the complete set of commutation relations
reads as follows:

\begin{eqnarray}
\left[e_a^{A A^{\prime}}\left(x\right),e_g^{G G^{\prime}}\left(y\right)\right]_{-}&=&\Gamma^{ceik}_{afgl}
\delta_{ed}\delta^{bf}\delta^{d}_c \delta_{kj}\delta^{hl}\delta^{j}_i 
\left[\delta_{bm}\delta^{np}\left(\mathcal{Q}^{-1}\right)^{mq}_{np}i e_{q}^{AA^{\prime}}\left(x\right)\right]
\left[\delta_{hr}\delta^{st}\left(\mathcal{Q}^{-1}\right)^{ru}_{st}i e_{u}^{GG^{\prime}}\left(y\right)\right]
,\nonumber\\
\left[e_a^{A A^{\prime}}\left(x\right),p^b_{B B^{\prime}}\left(y\right)\right]_{-}&=&
i\delta_{ed}\delta^{bf}\mathcal{Q}^{ce}_{af}\epsilon^A_{\ B} \epsilon^{A^{\prime}}_{\ B^{\prime}}\delta^d_c \delta(x-y),
\nonumber\\
\left[p^a_{A A^{\prime}}\left(x\right),p^i_{I I^{\prime}}\left(y\right)\right]_{-}&=&
\Gamma^{cekm}_{bfjn}\delta_{ed}\delta^{af}\delta^{d}_c\delta_{ml}\delta^{in}\delta^{l}_k 
\nonumber\\&&
\times \left\{\delta^{uv}
\left(\mathcal{Q}^{-1}\right)^{wb}_{uv}\left[ip_{AA^{\prime}w}\left(x\right)
-\frac{1}{2}\delta_{sr}\mathcal{Q}^{st}_{wo}\epsilon^{rqp}\psi_{A q}\left(x\right)D_{A^{\prime}t p}^B\left(x\right)
\left(\mathcal{C}_{BC^{\prime}}^{xy}\left(x\right)\left(\mathcal{Q}^{-1}\right)^{oz}_{xy}i\bar \psi_z^{C^{\prime}}\left(x\right)\right)
\right]\right\}
\nonumber\\&&
\times \left\{\delta^{\bar u \bar v}
\left(\mathcal{Q}^{-1}\right)^{\bar w j}_{\bar u\bar v}\left[ip_{II^{\prime}\bar w}\left(y\right)
-\frac{1}{2}\delta_{\bar s \bar r}\mathcal{Q}^{\bar s \bar t}_{\bar w \bar o}
\epsilon^{\bar r \bar q \bar p}\psi_{I \bar q}\left(y\right)D_{I^{\prime}\bar t \bar p}^D\left(y\right)
\left(\mathcal{C}_{DE^{\prime}}^{\bar x \bar y}\left(y\right)
\left(\mathcal{Q}^{-1}\right)^{\bar o \bar z}_{\bar x \bar y}
i\bar \psi_{\bar z}^{E^{\prime}}\left(y\right)\right)
\right]\right\}
\nonumber\\&&
+\frac{1}{2}\Gamma^{cenq}_{bfmj}
\delta_{ed}\delta^{af}\delta^{d}_c\delta_{pn}\delta^{im}
\epsilon^{plk}\psi_{Il}\left(y\right) D^{J}_{I^{\prime} qk}\left(y\right)
\left\{\delta^{\bar h \bar i}
\left(\mathcal{Q}^{-1}\right)^{\bar j b}_{\bar h \bar i}\left[ip_{AA^{\prime}\bar j}\left(x\right)
\right.\right.\nonumber\\&&\left.\left.
\quad-\frac{1}{2}\delta_{\bar f \bar e}\mathcal{Q}^{\bar f \bar g}_{\bar j \bar b}\epsilon^{\bar e \bar d \bar c}
\psi_{A \bar d}\left(x\right)D_{A^{\prime}\bar g \bar c}^B\left(x\right)
\left(\mathcal{C}_{BC^{\prime}}^{\bar k \bar l}\left(x\right)
\left(\mathcal{Q}^{-1}\right)^{\bar b \bar m}_{\bar k \bar l}
i\bar \psi_{\bar m}^{C^{\prime}}\left(x\right)\right)
\right]\right\}
\left[\mathcal{C}_{JB^{\prime}}^{yz}\left(y\right)\left(\mathcal{Q}^{-1}\right)^{jx}_{yz}i\bar \psi_x^{B^{\prime}}\left(y\right)\right]
\nonumber\\&&
+\frac{1}{2}\Gamma^{fhkm}_{ebjn}
\delta_{gf}\delta^{ae}\epsilon^{gdc}\delta_{ml}\delta^{in}\delta^{l}_k \psi_{Ad}\left(x\right)
D^{B}_{A^{\prime} hc}\left(x\right)
\left[\mathcal{C}_{BB^{\prime}}^{yz}\left(x\right)\left(\mathcal{Q}^{-1}\right)^{bx}_{yz}i\bar \psi_x^{B^{\prime}}\left(x\right)\right]
\nonumber\\&&
\quad\times \left\{\delta^{\bar h \bar i}
\left(\mathcal{Q}^{-1}\right)^{\bar jj}_{\bar h \bar i}\left[ip_{II^{\prime}\bar j}\left(y\right)
-\frac{1}{2}\delta_{\bar f \bar e}\mathcal{Q}^{\bar f \bar g}_{\bar j \bar b}\epsilon^{\bar e \bar d \bar c}
\psi_{A \bar d}\left(y\right)D_{A^{\prime} \bar g \bar c}^C\left(y\right)
\left(\mathcal{C}_{CC^{\prime}}^{\bar k \bar l}\left(y\right)
\left(\mathcal{Q}^{-1}\right)^{\bar b \bar m}_{\bar k \bar l}
i\bar \psi_{\bar m}^{C^{\prime}}\left(y\right)\right)
\right]\right\}
\nonumber\\&&
+\frac{1}{4}\Gamma^{fhnq}_{ebmj}
\delta_{gf}\delta^{ae}\epsilon^{gdc}\delta_{pn}\delta^{im}\epsilon^{plk}
\psi_{Ad}\left(x\right)D^{B}_{A^{\prime} hc}\left(x\right) \psi_{Il}\left(y\right) D^{J}_{I^{\prime} qk}\left(y\right)
\nonumber\\&&
\quad \times \left[\mathcal{C}_{BB^{\prime}}^{vw}\left(x\right)\left(\mathcal{Q}^{-1}\right)^{bu}_{vw}i\bar \psi_u^{B^{\prime}}\left(x\right)\right]
\left[\mathcal{C}_{JC^{\prime}}^{yz}\left(y\right)\left(\mathcal{Q}^{-1}\right)^{jx}_{yz}i\bar \psi_x^{C^{\prime}}\left(y\right)\right]
\nonumber\\&&
+\frac{1}{4}i\mathcal{Q}^{fh}_{eb}i\mathcal{Q}^{nq}_{mj}
\delta_{gf}\delta^{ae}\epsilon^{gdc}\delta_{pn}\delta^{im}\epsilon^{plk}
\left[\psi_{Ad}\left(x\right)D^{B}_{A^{\prime} hc}\left(x\right),
\psi_{Il}\left(y\right) D^{J}_{I^{\prime} qk}\left(y\right)\right]
\nonumber\\&&
\quad \times \left[\mathcal{C}_{BB^{\prime}}^{vw}\left(x\right)\left(\mathcal{Q}^{-1}\right)^{bu}_{vw}i\bar \psi_u^{B^{\prime}}\left(x\right)\right]
\left[\mathcal{C}_{JC^{\prime}}^{yz}\left(y\right)\left(\mathcal{Q}^{-1}\right)^{jx}_{yz}i\bar \psi_x^{C^{\prime}}\left(y\right)\right],
\nonumber\\
\left[\psi_a^A(x),\psi_e^E(y)\right]_{+}&=&
\Gamma^{cdgh}_{bafe}
D_{cd}^{A A^{\prime}}\left(x\right)
D_{gh}^{E E^{\prime}}\left(y\right)
\left[\mathcal{C}_{BA^{\prime}}^{jk}\left(x\right)\left(\mathcal{Q}^{-1}\right)^{bi}_{jk}i\psi_i^{B}\left(x\right)\right]
\left[\mathcal{C}_{CE^{\prime}}^{mn}\left(y\right)\left(\mathcal{Q}^{-1}\right)^{fl}_{mn}i\psi_l^{C}\left(y\right)\right]
\nonumber\\&&
+i\mathcal{Q}^{cd}_{ba}i\mathcal{Q}^{gh}_{fe}
\left[D_{cd}^{A A^{\prime}}\left(x\right),D_{gh}^{E E^{\prime}}\left(y\right)\right]
\left[\mathcal{C}_{BA^{\prime}}^{jk}\left(x\right)\left(\mathcal{Q}^{-1}\right)^{bi}_{jk}i\psi_i^{B}\left(x\right)\right]
\left[\mathcal{C}_{CE^{\prime}}^{mn}\left(y\right)\left(\mathcal{Q}^{-1}\right)^{fl}_{mn}i\psi_l^{C}\left(y\right)\right]
,\nonumber\\
\left[\psi_a^A\left(x\right),\bar \psi_b^{A^{\prime}}\left(y\right)\right]_{+}&=&
-i\mathcal{Q}^{cd}_{ab}D_{cd}^{A A^{\prime}}\left(x\right)\delta(x-y)
,\nonumber\\
\left[e_a^{A A^{\prime}}\left(x\right),\psi_g^{G}\left(y\right)\right]_{-}&=&
\Gamma^{ceij}_{afhg}\delta_{ed}\delta^{bf}\delta^{d}_c D_{ij}^{G G^{\prime}}\left(y\right)
\left[\delta_{bk}\delta^{lm}\left(\mathcal{Q}^{-1}\right)^{kn}_{lm}i e_{n}^{AA^{\prime}}\left(x\right)\right]
\left[\mathcal{C}_{BG^{\prime}}^{yz}\left(y\right)\left(\mathcal{Q}^{-1}\right)^{hx}_{yz}i\psi_x^{B}\left(y\right)\right]
,\nonumber\\
\left[p^a_{A A^{\prime}}\left(x\right),\psi_b^B\left(y\right)\right]_{-}&=&
\frac{i}{2}\delta_{gf}\delta^{ae}\mathcal{Q}^{fh}_{eb}
\epsilon^{gcd}\psi_{Ad}\left(x\right)D^{B}_{A^{\prime} hc}\left(x\right)\delta(x-y),
\label{generalized_quantization_supergravity_B}
\end{eqnarray}
where have been used the following relations:

\begin{eqnarray}
\frac{\delta}{\delta e_a^{AA^{\prime}}}&=&\delta^{hi}
\left(\mathcal{Q}^{-1}\right)^{ja}_{hi}\left[ip_{AA^{\prime}j}
-\frac{1}{2}\delta_{fe}\mathcal{Q}^{fg}_{jb}\epsilon^{edc}\psi_{A d}D_{A^{\prime}g c}^B
\left(\mathcal{C}_{BC^{\prime}}^{kl}\left(\mathcal{Q}^{-1}\right)^{bm}_{kl}i\bar \psi_m^{C^{\prime}}\right)
\right],\nonumber\\
\frac{\delta}{\delta p^a_{AA^{\prime}}}&=&-\delta_{ab}\delta^{cd}\left(\mathcal{Q}^{-1}\right)^{be}_{cd}i e_{e}^{AA^{\prime}},\quad
\frac{\delta}{\delta \psi_a^A}=\mathcal{C}_{AB^{\prime}}^{cd}
\left(\mathcal{Q}^{-1}\right)^{ab}_{cd}i\bar \psi_b^{B^{\prime}},\quad
\frac{\delta}{\delta \bar \psi_a^{A^{\prime}}}=\mathcal{C}_{BA^{\prime}}^{cd}
\left(\mathcal{Q}^{-1}\right)^{ab}_{cd}i\psi_b^{B},
\end{eqnarray}
where $\mathcal{C}_{AA^{\prime}}^{ab}$ as inverse of $D^{AA^{\prime}}_{ab}$ is defined according to

\begin{equation}
\mathcal{C}_{AA^{\prime}}^{ab}=-\frac{i\sqrt{h}}{2} e^{BB^{\prime} a} e_{AB^{\prime}}^b\ n_{B A^{\prime}}.
\label{definition_C}
\end{equation}
$\mathcal{C}_{AA^{\prime}}^{ab}$ is the inverse of $D^{AA^{\prime}}_{ab}$ in the sense that the following
relation is valid:

\begin{equation}
D^{AA^{\prime} a}_a \mathcal{C}^b_{BA^{\prime} b}=\epsilon^{A}_B,
\label{relation_inverse}
\end{equation}
where the contracted quantities $D_{a}^{A A^{\prime} a}$ and $\mathcal{C}_{A A^{\prime} a}^a$ appear,
which are obtained from ($\ref{definition_D}$) and ($\ref{definition_C}$),
$D_{a}^{A A^{\prime} a}=-\frac{2i}{\sqrt{h}} n^{AA^{\prime}}$,
$\mathcal{C}_{A A^{\prime} a}^a=-\frac{i\sqrt{h}}{2}n_{A A^{\prime}}$.
In the following calculation the validity of the relation ($\ref{relation_inverse}$) can be shown,

\begin{eqnarray}
D^{AA^{\prime} a}_a \mathcal{C}^b_{BA^{\prime} b}&=&\left[\frac{2i}{\sqrt{h}} n^{AA^{\prime}}\right]
\left[\frac{i\sqrt{h}}{2}n_{BA^{\prime}}\right]
=-n^{AA^{\prime}}n_{BA^{\prime}}=-n^\mu n^\nu \sigma_\mu^{AA^{\prime}} \sigma_{\nu BA^{\prime}}
=-n^\mu n^\nu \left(\delta_{\mu\nu}\epsilon^A_B+i\epsilon_{\mu\nu\rho}\sigma_{\rho B}^{A}\right)\nonumber\\
&=&-n^\mu n_\mu \epsilon^{A}_B=\epsilon^{A}_B,\quad {\rm since}\quad \epsilon_{\mu\nu\rho}=-\epsilon_{\nu\mu\rho}
\quad {\rm and}\quad n^\mu n_\mu=-1.
\label{proof_relation_inverse}
\end{eqnarray}
The relation ($\ref{relation_inverse}$) is also very important concerning the definition of the inner product
defining the corresponding Hilbert space. The above constraints of classical $\mathcal{N}=1$ supergravity
($\ref{constraints}$) lead to the following quantum constraints:

\begin{equation}
\hat {\mathcal{H}}_\tau|\Psi \rangle=0,\quad \hat {\mathcal{H}}_a |\Psi \rangle=0,\quad
\hat S_A |\Psi \rangle=0,\quad {\hat {\bar S}}_{A^{\prime}}|\Psi \rangle=0,\quad
\hat J_{AB}|\Psi \rangle=0,\quad {\hat {\bar J}}_{A^{\prime}B^{\prime}}|\Psi \rangle=0.
\label{quantum_constraints_supergravity}
\end{equation}
If these quantum constraints ($\ref{quantum_constraints_supergravity}$) are written explicitly and the
representation referring to the states expressed by $e_a^{AA^{\prime}}$ and $\psi_a^{AA^{\prime}}$, $|\Psi\left[e,\psi\right]\rangle$ ($\ref{operators_A1}$),
($\ref{operators_A2}$), is considered, then they read as follows:

\begin{eqnarray}
\hat {\mathcal{H}}_\tau|\Psi \rangle&=&\left[16 \pi G G_{abcd}\hat {\pi}^{ab}\hat {\pi}^{cd}
-\frac{\sqrt{\hat h}\hat R}{16 \pi G}
+\left(\frac{1}{2}\epsilon^{abc}\hat {\bar \psi}_a^{A^{\prime}}n_{AA^{\prime}}
\mathcal{D}_b \hat \psi_c^A+h.c.\right)\right]|\Psi \rangle\nonumber\\
&=&\left\{16 \pi G G_{abcd}
\frac{e^{AA^{\prime}(a}
\left[-i\delta_{hg}\delta^{b)i}\mathcal{Q}^{fh}_{ei}\delta^{g}_f \frac{\delta}{\delta e_e^{A A^{\prime}}}
-\frac{i}{2}\delta_{ji}\delta^{b)h}\mathcal{Q}^{ik}_{he}\epsilon^{jgf}\psi_{Ag}D^{B}_{A^{\prime} kf}
\frac{\delta}{\delta \psi^B_e}\right]}{2}
\right.\nonumber\\&&\left.
\cdot \frac{e^{BB^{\prime}(c}
\left[-i\delta_{hg}\delta^{d)i}\mathcal{Q}^{fh}_{ei}\delta^{g}_f \frac{\delta}{\delta e_e^{A A^{\prime}}}
-\frac{i}{2}\delta_{ji}\delta^{d)h}\mathcal{Q}^{ik}_{he}\epsilon^{jgf}\psi_{Ag}D^{B}_{A^{\prime} kf}
\frac{\delta}{\delta \psi^B_e}\right]}{2}
\right.\nonumber\\&&\left.
-\frac{\sqrt{h} R}{16 \pi G}
+\left[\frac{1}{2}\epsilon^{abc}
 \left(-i\mathcal{Q}^{gh}_{fa}D_{gh}^{A^{\prime} B}\frac{\delta}{\delta \psi_f^{B}}\right)n_{AA^{\prime}}
\mathcal{D}_b \psi_c^A+h.c.\right]\right\}|\Psi\left(e,\psi\right)\rangle=0,
\end{eqnarray}

\begin{eqnarray}
\hat {\mathcal{H}}_a|\Psi \rangle&=&\left[-2 \hat h_{ab} \mathcal{D}_c {\hat \pi}^{bc}
+\left(\frac{1}{2}\epsilon^{bcd}\hat {\bar \psi}_b^{A^{\prime}}\hat e_{AA^{\prime} a}\mathcal{D}_c {\hat \psi}_d^A
+h.c.\right)\right]|\Psi \rangle\nonumber\\
&=&\left\{-2 \hat h_{ab} \mathcal{D}_c
\frac{\hat e^{AA^{\prime}(b}
\left[-i\delta_{gf}\delta^{c)h}\mathcal{Q}^{eg}_{dh}\delta^{f}_e \frac{\delta}{\delta e_d^{A A^{\prime}}}
-\frac{i}{2}\delta_{ih}\delta^{c)g}\mathcal{Q}^{hj}_{gd}\epsilon^{ife}\psi_{Af}D^{B}_{A^{\prime} je}
\frac{\delta}{\delta \psi^B_d}\right]}{2}
\right.\nonumber\\&&\left.
+\left[\frac{1}{2}\epsilon^{bcd}
\left(-i\mathcal{Q}^{gh}_{fb}D_{gh}^{A^{\prime} B}\frac{\delta}{\delta \psi_f^{B}}\right)
\hat e_{AA^{\prime} a}\mathcal{D}_c {\hat \psi}_d^A
+h.c.\right]\right\}|\Psi\left(e,\psi\right)\rangle=0,
\end{eqnarray}

\begin{eqnarray}
\hat S_A|\Psi \rangle&=&\left[\epsilon^{abc}\hat e_{A A^{\prime} a}\mathcal{D}_b \hat {\bar \psi}^{A^{\prime}}_c
-i4\pi G \hat p_{A A^{\prime}}^a \hat {\bar \psi}^{A^{\prime}}_a\right]|\Psi \rangle\nonumber\\
&=&\left[\epsilon^{abc} e_{A A^{\prime} a}\mathcal{D}_b 
\left(-i\mathcal{Q}^{gh}_{fc}D_{gh}^{A^{\prime} B}\frac{\delta}{\delta \psi_f^{B}}\right)
-i4\pi G \left(-i\delta_{ed}\delta^{af}\mathcal{Q}^{ce}_{bf}\delta^{d}_c \frac{\delta}{\delta e_b^{A A^{\prime}}}
\right.\right.\nonumber\\&&\left.\left.
-\frac{i}{2}\delta_{gf}\delta^{ae}\mathcal{Q}^{fh}_{eb}\epsilon^{gdc}\psi_{Ad}D^{B}_{A^{\prime} hc}
\frac{\delta}{\delta \psi^B_b}\right)
\cdot\left(-i\mathcal{Q}^{gh}_{fa}D_{gh}^{A^{\prime} B}\frac{\delta}{\delta \psi_f^{B}}\right)
\right]|\Psi\left(e,\psi\right) \rangle=0,
\end{eqnarray}

\begin{eqnarray}
{\hat {\bar S}}_{A^{\prime}}|\Psi \rangle&=&\left[\epsilon^{abc} \hat e_{A A^{\prime} a}\mathcal{D}_b \hat \psi^{A}_c
+i4\pi G \hat \psi^{A^{\prime}}_a \hat p_{A A^{\prime}}^{a}\right]|\Psi \rangle\nonumber\\
&=&\left[\epsilon^{abc} e_{A A^{\prime} a}\mathcal{D}_b \psi^{A}_c
+i4\pi G \psi^{A^{\prime}}_a 
\left(-i\delta_{ed}\delta^{af}\mathcal{Q}^{ce}_{bf}\delta^{d}_c \frac{\delta}{\delta e_b^{A A^{\prime}}}
\right.\right.\nonumber\\&&\left.\left.
-\frac{i}{2}\delta_{gf}\delta^{ae}\mathcal{Q}^{fh}_{eb}\epsilon^{gdc}\psi_{Ad}D^{B}_{A^{\prime} hc}
\frac{\delta}{\delta \psi^B_b}
\right)
\right]|\Psi\left(e,\psi\right) \rangle=0,
\end{eqnarray}

\begin{eqnarray}
\hat J_{AB}|\Psi \rangle&=&\left[\hat e_{(A}^{A^{\prime} a}\hat p_{B) A^{\prime} a}
-\frac{1}{2}\hat \psi_{(A}^a \epsilon_{abc}\hat {\bar \psi}^{A^{\prime}b} \hat e_{B) A^{\prime}}^c\right]|\Psi \rangle
\nonumber\\
&=&\left[e_{(A}^{A^{\prime} a}
\left(
-i\delta_{ed}\mathcal{Q}^{ce}_{ba}\delta^{d}_c \frac{\delta}{\delta e_b^{B) A^{\prime}}}
-\frac{i}{2}\delta_{fe}\mathcal{Q}^{eg}_{ab}\epsilon^{fdc}\psi_{B)d}D^{C}_{A^{\prime} gc}
\frac{\delta}{\delta \psi^C_b}
\right)
\right.\nonumber\\&&\left.
-\frac{1}{2} \psi_{(A}^a \epsilon_{abc}\left(-i\delta^{be}\mathcal{Q}^{gh}_{fe}D_{cd}^{A^{\prime} C}
\frac{\delta}{\delta \psi_f^{C}}\right)e_{B) A^{\prime}}^c\right]|\Psi\left(e,\psi\right) \rangle=0,
\end{eqnarray}

\begin{eqnarray}
{\hat {\bar J}}_{A^{\prime}B^{\prime}}|\Psi \rangle&=&\left[\hat e_{(A^{\prime}}^{A a}\hat p_{B^{\prime}) A a}
-\frac{1}{2}\hat {\bar \psi}_{(A^{\prime}}^a \epsilon_{abc}\hat \psi^{A b} \hat e_{B^{\prime}) A}^c\right]
|\Psi \rangle\nonumber\\
&=&\left[e_{(A^{\prime}}^{A a}
\left(-i\delta_{ed}\mathcal{Q}^{ce}_{ba}\delta^{d}_c \frac{\delta}{\delta e_b^{B^{\prime}) A}}
-\frac{i}{2}\delta_{fe}\mathcal{Q}^{eg}_{ab}\epsilon^{fdc}\psi_{B^{\prime})d}D^{C}_{A gc}
\frac{\delta}{\delta \psi^C_b}
\right)
\right.\nonumber\\&&\left.
-\frac{1}{2}\epsilon_{(A^{\prime}D^{\prime}}
\left(-i\delta^{ae}\mathcal{Q}^{gh}_{fe}D_{gh}^{D^{\prime} C}
\frac{\delta}{\delta \psi_f^{C}}\right)
\epsilon_{abc} \psi^{A b} e_{B^{\prime}) A}^c\right]
|\Psi\left(e,\psi\right) \rangle=0.
\end{eqnarray}
To define the Hilbert space of the states, $|\Psi\left[e,\psi\right]\rangle$
or $|\Psi\left[p,\bar \psi\right]\rangle$ respectively, of the quantum
theory of $\mathcal{N}=1$ supergravity based on the quaternionic quantization principle,
$\mathcal{H}_{\mathcal{N}=1}$, an inner product has to be specified finally.
The inner product $\langle\ \cdot\ |\ \cdot \ \rangle$ between two quantum states of quantum
supergravity under presupposition of the quaternionic quantization principle is
formulated as follows:

\begin{equation}
\langle \Phi|\Psi \rangle=\int \mathcal{D}e \mathcal{D}\psi \mathcal{D}\bar \psi\ \bar \Phi\left(e,\bar \psi\right)
\Psi\left(e,\psi\right) \exp\left[-\int d^3 x \left(\mathcal{Q}^{-1}\right)^{ab}_{cd}i
\mathcal{C}_{AA^{\prime}}^{cd}(x) \psi^{A}_a(x) \bar \psi^{A^{\prime}}_b(x)\right]
D^{-1}\left(e\right),
\label{inner_product}
\end{equation}
where $D\left(e\right)$ is defined as

\begin{equation}
D\left(e\right)=\prod_x \det\left[-i \mathcal{C}_{AA^{\prime}}^{ab}(x)\right].
\end{equation}
The definition of the inner product according to ($\ref{inner_product}$) represents the corresponding generalization
of the inner product given in \cite{D'Eath:1984sh} to the case of the quaternionic quantization principle. Besides
incorporating $\left(\mathcal{Q}^{-1}\right)^{ab}_{cd}$ to the inner product, the quantity $C_{AA^{\prime}}^{ab}=
-\epsilon^{abc}e_{AA^{\prime}c}(x)$ appearing in the usual formulation of the inner product of quantum supergravity
has been replaced by $\mathcal{C}_{AA^{\prime}}^{ab}$, which has already been defined in ($\ref{definition_C}$).
The definition of the inner product according to ($\ref{inner_product}$) maintains that the operators
$\hat \psi^{A}_a$ and $\hat {\bar \psi}^{A^{\prime}}_a$ are still hermitian conjugated quantities to
each other in this generalized quantization scenario,

\begin{equation}
\langle \Phi|\hat {\bar \psi}^{A^{\prime}}_a|\Psi\rangle=\langle \hat \psi^{A}_a\Phi|\Psi \rangle.
\end{equation}
This can be shown as follows:

\begin{eqnarray}
\langle \Phi|\hat {\bar \psi}^{A^{\prime}}_a|\Psi \rangle
&=&\int \mathcal{D}e \mathcal{D}\psi \mathcal{D}\bar \psi\ \bar \Phi\left(e,\bar \psi\right)
\left[-i\mathcal{Q}^{cd}_{ba}D_{cd}^{A A^{\prime}}\frac{\delta \Psi\left(e,\psi\right)}
{\delta \psi_b^{A}}\right]
\exp\left[-\int d^3 x \left(\mathcal{Q}^{-1}\right)^{ef}_{gh}i
\mathcal{C}_{BB^{\prime}}^{cd}(x) \psi^{B}_e(x) \bar
\psi^{B^{\prime}}_f(x)\right]D^{-1}\left(e\right),\nonumber\\
&=&\int \mathcal{D}e \mathcal{D}\psi \mathcal{D}\bar \psi\
i\mathcal{Q}^{cd}_{ba}D_{cd}^{A A^{\prime}}\frac{\delta}{\delta \psi_b^{A}}
\left[\Phi\left(e,\bar \psi\right)\exp\left[-\int d^3 x \left(\mathcal{Q}^{-1}\right)^{ef}_{gh}i
\mathcal{C}_{BB^{\prime}}^{cd}(x) \psi^{B}_e(x) \bar
\psi^{B^{\prime}}_f(x)\right]D^{-1}\left(e\right)\right]
\Psi\left(e,\psi\right),\nonumber\\
&=&\int \mathcal{D}e \mathcal{D}\psi \mathcal{D}\bar \psi\
i\mathcal{Q}^{cd}_{ba}D_{cd}^{A A^{\prime}}\frac{\delta}{\delta \psi_b^{A}}
\left[-\int d^3 x \left(\mathcal{Q}^{-1}\right)^{ef}_{gh}i
\mathcal{C}_{BB^{\prime}}^{cd}(x) \psi^{B}_e(x) \bar
\psi^{B^{\prime}}_f(x)\right]
\nonumber\\
&&\quad \times \Phi\left(e,\bar \psi\right)\exp\left[-\int d^3 x \left(\mathcal{Q}^{-1}\right)^{ef}_{gh}i
\mathcal{C}_{BB^{\prime}}^{cd}(x) \psi^{B}_e(x) \bar
\psi^{B^{\prime}}_f(x)\right]D^{-1}\left(e\right)
\Psi\left(e,\psi\right),\nonumber\\
&=&\int \mathcal{D}e \mathcal{D}\psi \mathcal{D}\bar \psi\ \bar \Phi\left(e,\bar \psi\right)
\bar \psi_a^{A^{\prime}}
\exp\left[-\int d^3 x \left(\mathcal{Q}^{-1}\right)^{ef}_{gh}i
\mathcal{C}_{BB^{\prime}}^{cd}(x) \psi^{B}_e(x) \bar
\psi^{B^{\prime}}_f(x)\right]D^{-1}\left(e\right),\nonumber\\
&=&\int \mathcal{D}e \mathcal{D}\psi \mathcal{D}\bar \psi\
\hat {\bar \psi}_a^{A^{\prime}} \bar \Phi\left(e,\bar \psi\right)
\exp\left[-\int d^3 x \left(\mathcal{Q}^{-1}\right)^{ef}_{gh}i
\mathcal{C}_{BB^{\prime}}^{cd}(x) \psi^{B}_e(x) \bar
\psi^{B^{\prime}}_f(x)\right]D^{-1}\left(e\right)
=\langle \hat \psi^{A}_a\Phi|\Psi \rangle,
\end{eqnarray}
where has been used partial integration in the second step including the boundary condition

\begin{eqnarray}
-\int \mathcal{D}e \mathcal{D}\bar \psi\
i\mathcal{Q}^{cd}_{ba}D_{cd}^{A A^{\prime}}
\left[\Phi\left(e,\bar \psi\right)\exp\left[-\int d^3 x \left(\mathcal{Q}^{-1}\right)^{ef}_{gh}i
\mathcal{C}_{BB^{\prime}}^{cd}(x) \psi^{B}_e(x) \bar
\psi^{B^{\prime}}_f(x)\right]D^{-1}\left(e\right)
\Psi\left(e,\psi\right)\right|^{\infty}_{-\infty}=0,
\end{eqnarray}
and the representation of the operator $\hat \psi^{A}_a$ with respect to the states depending on $e$ and
$\bar \psi$, $|\Psi\left(e,\bar \psi\right)\rangle$, which has been given in ($\ref{operators_B1}$).
In the fourth step has been performed the following relation:

\begin{equation}
i\mathcal{Q}^{cd}_{ba}D_{cd}^{A A^{\prime}}\frac{\delta}{\delta \bar \psi_b^{A^{\prime}}\left(x\right)}
\left[-\int d^3 x^{\prime} \left(\mathcal{Q}^{-1}\right)^{ef}_{gh}
i\mathcal{C}_{BB^{\prime}}^{gh}(x^{\prime})\psi^{B}_e(x^{\prime}) \bar \psi^{B^{\prime}}_f(x^{\prime})\right]
=\psi^A_a\left(x\right)\left[-\int d^3 x^{\prime} \left(\mathcal{Q}^{-1}\right)^{ef}_{gh}
i\mathcal{C}_{BB^{\prime}}^{gh}(x^{\prime})\psi^{B}_e(x^{\prime})\bar \psi^{B^{\prime}}_f(x^{\prime})\right],
\end{equation}
which is based on the calculation

\begin{eqnarray}
&&-i\mathcal{Q}^{cd}_{ba}D_{cd}^{A A^{\prime}}\left(\mathcal{Q}^{-1}\right)^{eb}_{gh}i\mathcal{C}_{BA^{\prime}}^{gh}\psi^{B}_e
=-\left[i\frac{\delta^{cd}}{3}\mathcal{P}_{ab}\right]\left[3\delta_{gh}\left(\mathcal{P}^{-1}\right)^{eb}i\right]
D_{cd}^{A A^{\prime}}\mathcal{C}_{BA^{\prime}}^{gh}\psi^{B}_e%\nonumber\\&&
=\delta^e_a D_{c}^{A A^{\prime} c}\mathcal{C}_{BA^{\prime} g}^g\psi^{B}_e
=\delta^e_a \epsilon^A_B \psi^B_e=\psi^A_a,\nonumber\\
\end{eqnarray}
where has been used that $\mathcal{P}^{ab}\left(\mathcal{P}^{-1}\right)_{bc}=\delta^a_c$ as well
as the relation ($\ref{relation_inverse}$), whose validity has been shown in
($\ref{proof_relation_inverse}$).
Also the Fourier transformations between the representation $|\Psi\left(e,\psi\right)\rangle$ and
$|\tilde \Psi\left(e,\bar \psi\right)\rangle$, which are wave-functionals, is defined by the inner
product ($\ref{inner_product}$), and the Fourier transformation corresponding to
($\ref{inner_product}$) reads as follows:

\begin{eqnarray}
&&\tilde \Psi\left(e,\bar \psi\right)=D^{-1}\left(e\right)\int \mathcal{D}\psi\ \Psi\left(e,\psi\right)
\exp\left[-\int d^3 x \left(\mathcal{Q}^{-1}\right)^{ab}_{cd}i\mathcal{C}_{AA^{\prime}}^{cd}(x)\psi^{A}_a(x)
\bar \psi^{A^{\prime}}_b(x)\right].
\end{eqnarray}
%and the corresponding inverse Fourier transformation accordingly reads

%\begin{equation}
%\Psi\left(e,\psi\right)=\int \mathcal{D}\psi\ \tilde \Psi\left(e,\bar \psi\right)
%\exp\left[\int d^3 x \left(\mathcal{Q}^{-1}\right)^{ab}_{cd}i\mathcal{C}_{AA^{\prime}}^{cd}(x)\psi^{A}_a(x)
%\bar \psi^{A^{\prime}}_b(x)\right].
%\end{equation}

\section{Summary and Discussion}

In this paper has been presented a generalized quantization principle for quantum theory based on an algebra
containing quaternions, which has been applied to quantum mechanics as well as to canonical quantum
general relativity and to $\mathcal{N}$=1 canonical quantum supergravity. The presented theory can be seen as an
alternative approach to formulate general quantum theory and especially to quantize general relativity.
The quaternionic quantization principle assumes that the components of a variable belonging to a special space-time
direction do not only fulfil nontrivial commutation relations with the corresponding component of the
canonical conjugated variable, but also with the other components of this variable. The additional
commutation relations are assumed to be proportional to another direction than the usual complex
direction in the space of quaternions. This implies also additional commutation relations between the
several components of the special variable with each other, leading to noncommutative geometry in
case of quantum mechanics and leading to commutation relations between the components of the gravitational
field in quantum gravity. This has its origin in the fact that the components of quaternions do not commute
with each other and accordingly also the components of the generalized quantization tensor, which is
contained in the generalized expressions of the operators, do not commute with each other.

Especially the application of this quantization principle to supergravity contains special intricacies,
which are related to the necessity to use Dirac brackets instead of Poisson brackets to
obtain the corresponding quantum theory from the classical theory. Accordingly the quaternionic
quantization principle has to be adapted to the special conditions of supergravity, what
has been done for the special case of $\mathcal{N}=1$ supergravity in this paper.
Because of the appearance of Dirac brackets in case of supergravity additional commutation relations
to the usual commutation relations based on the Heisenberg algebra have to be considered, which are
defined by the commutation relations between the several operators and are decisively extended in case
of a presupposition of the quaternionic quantization principle. Besides, the inner product of canonical
quantum supergravity has to be modified to maintain that the conjugated quantity to the Rarita Schwinger
field is still the hermitian adjoint quantity.

The relation between this generalization of the quantization principle of general quantum theory to
the usual one is determined by the new dimensionless parameter $\varkappa$. If $\varkappa$ goes to
zero, the generalized theory is approximatively equal to the usual quantum theory.  
In principle, it is also thinkable that the quaternionic quantization principle is just valid for
the special case of the quantization of gravity, as quantization of usual general relativity or
supergravity respectively. In this case quantum mechanics and quantum field theory would not have
to be changed, but quantum gravity would be based on such a generalized quantization principle.

The decisive intention of the paper consists in the assumption that the formulation of a quantum
theory of general relativity could not only presuppose a generalization of the classical approximation
or at least its formulation, but could alternatively or additionally presuppose a generalization of
the quantization principle as way to obtain a quantum theory from a classical theory.
The presented idea of such a generalization of the concept of quantization can be interpreted
as a special manifestation of this principle consideration, which is based on a suggesting
generalization, because there is no reason why quantum theory should be restricted to complex
Hilbert spaces and why the commutation relations between the components of the variables and the
canonical conjugated variables belonging to different space-time directions should be assumed
to commute. This means that if the possibility of a generalization of quantum theory as it is
already postulated in noncommutative geometry with respect to the several components of
space-time coordinates is assumed to make sense, the presented theory appears as a very promising
candidate for such a generalization, which can in principle also be transferred to other
formulations of general relativity as the loop representation or to extended
supergravity theories.

\end{document}